\documentclass[12pt,preprint]{aastex}

\usepackage{graphicx}

\newcommand{\e}{e}
\newcommand{\degrees}{^\circ}
\newcommand{\imag}{i}

\shorttitle{A Map of the Universe}
\shortauthors{Gott, Juric et al.}

\begin{document}

\title{A Map of the Universe}

\author{
J. Richard Gott, III, \altaffilmark{1}
Mario Juri\'c, \altaffilmark{1}
David Schlegel, \altaffilmark{1}
Fiona Hoyle, \altaffilmark{2}
Michael Vogeley, \altaffilmark{2}
Max Tegmark, \altaffilmark{3}
Neta Bahcall, \altaffilmark{1}
Jon Brinkmann \altaffilmark{4}
}

\altaffiltext{1}{Department of Astrophysical Sciences, Princeton University, Princeton, NJ 08544 }
\altaffiltext{2}{Department of Physics, Drexel University, Philadelphia, PA 19104}
\altaffiltext{3}{Department of Physics, University of Pennsylvania, Philadelphia, PA 19104}
\altaffiltext{4}{Apache Point Observatory, 2001 Apache Point Road, P.O Box 59,Sunspot, NM, 88349}

\begin{abstract}
We have produced a new conformal map of the universe illustrating recent
discoveries, ranging from Kuiper belt objects in the Solar system, to the
galaxies and quasars from the Sloan Digital Sky Survey.  This map
projection, based on the logarithm map of the complex plane, preserves
shapes locally, and yet is able to display the entire range of astronomical
scales from the Earth's neighborhood to the cosmic microwave background. 
The conformal nature of the projection, preserving shapes locally, may be of
particular use for analyzing large scale structure. Prominent in the map is
a Sloan Great Wall of galaxies 1.37 billion light years long, 80\%
longer than the Great Wall discovered by Geller and Huchra and therefore the
largest observed structure in the universe.
\end{abstract}


\section{Introduction}

Cartographers mapping the Earth's surface were faced with the challenge of
mapping a curved surface onto a plane.  No such projection can be perfect,
but it can capture important features.  Perhaps the most famous map
projection is the Mercator projection (presented by Gerhardus Mercator in
1569).  This is a conformal projection which preserves shapes locally. 
Lines of latitude are shown as straight horizontal lines, while meridians of
longitude are shown as straight vertical lines.  If the Mercator projection
is plotted on an $(x,y)$ plane, the coordinates are plotted as follows: $x =
\lambda$, and $y = \ln(\tan(\pi/4 + \phi/2))$ where $\phi$
(positive if north, negative if south) is the latitude in radians, while
$\lambda$ (positive if easterly, negative if westerly) is the longitude in
radians (see \cite{sny93} for an excellent discussion of this and other map
projections of the Earth.) This conformal map projection preserves angles
locally, and also compass directions.  Local shapes are good, while the
scale varies as a function of latitude.  Thus, the shapes of both Iceland
and South America are shown well, although Iceland is shown larger than it
should be relative to South America. Other map projections preserve other
properties. The stereographic projection which, like the Mercator
projection, is conformal is often used to map hemispheres.  The gnomonic map
projection (effectively from a "light" at the center of the globe onto a
tangent plane) maps geodesics into straight lines on the flat map, but does
not preserve shapes or areas.  Equal area map projections like the Lambert,
Mollweide, and Hammer projections preserve areas but not shapes.

A Lambert azimuthal equal area projection, centered on the north pole has in
polar coordinates ($r$, $\theta$), $\theta = \lambda$, $r = 2 r_0 \sin
[(\pi/2 - \phi) / 2]$, where $r_0$ is the radius of the sphere. This
projection preserves areas. The northern hemisphere is thus mapped onto a
circular disk of radius $\sqrt{2}r_0$ and area $2 \pi r_0^2$. An oblique
version of this, centered at a point on the equator,  is also possible.

The Hammer projection shows the Earth as a horizontal ellipse with 2:1 axis
ratio. The equator is shown as a straight horizontal line marking the long
axis of the ellipse. It is produced in the following way: map the entire
sphere onto its western hemisphere by simply compressing each longitude by a
factor of 2. Now map this western hemisphere onto a plane by the Lambert
equal area azimuthal projection. This map is a circular disk. This is then
stretched by a factor of 2 (undoing the previous compression by a factor of
2) in the equatorial direction to make an ellipse with a 2:1 axis ratio.
Thus, the Hammer projection preserves areas. The Mollweide projection also
shows the sphere as 2:1 axis ratio ellipse. $(x, y)$ coordinates on the map
are: $x = (2 \sqrt{2}/\pi) r_0 \lambda \cos{\theta}$, and $y = \sqrt{2} r_0
\sin{\theta}$ where $2\theta + \sin 2 \theta = \pi \sin \phi$). This
projection is equal area as well. Latitude lines on the Mollweide projection
are straight, whereas they are curved arcs on the Hammer projection.

Astronomers mapping the sky have also used such map projections of the
sphere. Gnomonic maps of the celestial sphere onto a cube date from
1674. In recent times, \cite{tg76} used the stereographic map projection
to chart groups of galaxies (utilizing its property of mapping circles in
the sky onto circles on the map.) The COBE satellite map (\cite{cobemap}) of
the cosmic microwave background used an equal area map projection of the
celestial sphere onto a cube. The WMAP satellite (\cite{wmap}) mapped the
celestial sphere onto a rhombic dodecahedron using the Healpix equal area
map projection (\cite{healpix}).  Its results were displayed also on the
Mollweide map projection, showing the celestial sphere as an ellipse, which
was chosen for its equal area property, and the fact that lines of constant
galactic latitude are shown as straight lines.

\cite{lgh86} pioneered use of slice maps of the universe to make flat maps. 
They surveyed a slice of sky, $117\degrees$ long and $6\degrees$ wide, of constant
declination.  In 3D this slice had the geometry of a cone, and they
flattened this onto a plane. (A cone has zero Gaussian curvature and can
therefore be constructed from a piece of paper. A cone cut along a line and
flattened onto a plane looks like a pizza with a slice missing .) If the
cone is at declination $\delta$, the map in the plane will be 
$x = r \cos(\lambda \cos(\delta))$, $y = r \sin(\lambda \cos(\delta))$, where $\lambda$ is the
right ascension (in radians), and $r$ is the co-moving distance (as indicated
by the redshift of the object).  This will preserve shapes.  Many times a
$360\degrees$ slice is shown as a circle with the Earth in the center, where $x
= r \cos(\lambda)$, $y = r\sin(\lambda)$. If $r$ is measured in co-moving
distance, this will preserve shapes only if the universe is flat ($k = 0$),
and the slice is in the equatorial plane ($\delta = 0$), (if $\delta \neq 0$, structures
(such as voids) will appear lengthened in the direction tangential to the
line of sight by a factor of $1/\cos(\delta)$). This correction is
important for study of the Alcock-Paczynski effect, which says that
structures such as voids will not be shown in proper shape if we take simply
$r = z$ (\cite{ap79}).  In fact, \cite{ryd95} and \cite{rm96} have emphasized that this
shape distortion in redshift space can be used to test the cosmological
model in a large sample such as the Sloan Digital Sky Survey
(\cite{sdss1,sdss2,sdss3}).  If voids run
into each other, the walls will on average not have systematic peculiar
velocities and therefore voids should have approximately round shapes (a
proposition which can be checked in detail with N-body simulations).
Therefore, it is important to investigate map projections which will
preserve shapes locally. If one has the correct cosmological model, and
uses such a conformal map projection, isotropic features in the large scale
structure will appear isotropic on the map.

Astronomers mapping the universe are confronted with the challenge of
showing a wide variety of scales. What should a map of the universe show?
It should show locations of all the famous things in space: the Hubble Space
Telescope, the International Space Station, other satellites orbiting the
Earth, the van Allen radiation belts, the Moon, the Sun, planets, asteroids,
Kuiper belt objects, nearby stars such as $\alpha$~Centauri, and Sirius, stars
with planets such as 51 Peg, stars in our galaxy, famous black holes and
pulsars, the galactic center, Large and Small Magellanic Clouds, M31, famous
galaxies like M87, the Great Wall, famous quasars like 3C273 (\cite{sch63}) and the
gravitationally lensed quasar 0957 (\cite{kun97}), distant Sloan Digital Sky Survey
galaxies, and quasars, the most distant known quasar and galaxy and finally
the cosmic microwave background radiation.  This is quite a challenge.
Perhaps the first book to address this challenge was \underline{Cosmic View: The
Universe in 40 Jumps} by Kees Boeke published in 1957 (\cite{boe57}).  This brilliant book
started with a picture of a little girl shown at $1/10$th scale.  The next
picture showed the same little girl at $1/100$th scale who now could be seen
sitting in her school courtyard.  Each successive picture was plotted at ten
times smaller scale.  The $8$th picture, at a scale of $1/10^8$, shows the entire
Earth.  The $14$th picture, at a scale of $1/10^{14}$, shows the entire Solar
system.  The $18$th picture, at a scale of $1/10^{18}$, includes $\alpha$~Centauri,
The $22^\mathrm{nd}$ picture, at a scale of $1/10^{22}$ shows all of the Milky Way Galaxy. 
The $26^\mathrm{th}$ and last picture in the sequence shows galaxies out to
a distance of 750 million light years. A further sequence of pictures
labeled $0, -1, -2, ... -13$, starting with a life size picture of the
girl's hand, shows a sequence of microscopic views, each ten times larger in
size, ending with a view of the nucleus of a sodium atom at a scale of
$10^{13}/1$.  A modern version of this book, \underline{Powers of Ten} 
(\cite{powoftenbook}) by Phillip and
Phylis Morrison and the Office of Charles \& Ray Eames, is probably
familiar to most astronomers. This successfully addresses the scale
problem, but is an atlas of maps, not a single map.  How does one show the
entire observable universe in a single map?

The modern \underline{Powers of Ten} book described above is based on a movie, 
\underline{Powers of Ten} (\cite{powoften}), by Charles and Ray Eames which in turn was
inspired by Kees Boeke's book. The movie is arguably an even more brilliant
presentation than Kees Boeke's original book.  The camera starts with a
picture of a couple sitting on a picnic blanket in Chicago, and then the camera
moves outward, increasing its distance from them exponentially as a
function of time.  Thus, approximately every ten seconds, the view is from
ten times further away and corresponds to the next picture in the book.  The
movie gives one long continuous shot, which is breathtaking as it moves out. 
The movie is called \underline{Powers of Ten} (and recently, an IMAX version
of this idea has been made, called "Cosmic Voyage"), but it could equally
well be titled Powers of Two, or Powers of $\e$, because its exponential
change of scale with time, produces a reduction by a factor of two in
constant time intervals, and also a factor of $\e$ in constant time
intervals.  The time intervals between factors of $10$, factors of $2$ and
factors of $\e$ in the movie are related by the ratios
$\ln 10 : \ln 2 : 1$.  Still, this is not a single map which can be studied
all at once, or which can be hung on a wall.

We want to see the large scale structure of galaxy clustering but are also
interested in stars in our own galaxy and the Moon and planets.  Objects
close to us may be inconsequential in terms of the whole universe but they
are important to us.  It reminds one of the famous cartoon New Yorker cover
"View of the World from $9^\mathrm{th}$ Avenue" by Saul Steinberg of May 29, 1976
(\cite{ste76}).  It
humorously shows a New Yorker's view of the world.  The traffic, sidewalks
and buildings along $9^\mathrm{th}$ Avenue are visible in the foreground.  Behind is the
Hudson river, with New Jersey as a thin strip on the far bank.  Then at even
smaller scale is the rest of the United States with the Rocky mountains
sticking up like small hills.  In the background, but not much wider than
the Hudson River, is the entire Pacific ocean with China and Japan in the
distance.  This is, of course a parochial view, but it is just that kind of
view that we want of the universe.  We would like a single map that would
equally well show both interesting objects in the solar system, nearby
stars, galaxies in the Local Group, and large scale structure out to the
cosmic microwave background.

\section{Co-moving coordinates}

Our objective here is to produce a conformal map of the universe which will
show the wide range of scales encountered while still showing shapes that
are locally correct.

Consider the general Friedmann metrics:
\begin{eqnarray}
ds^2 & = & -dt^2 + a^2(t)(d\chi^2 + \sin^2\chi (d\theta^2 + \sin^2\theta{}d\phi^2)), \quad k = +1 \\
ds^2 & = & -dt^2 + a^2(t)(d\chi^2 + \chi^2(d\theta^2 + \sin^2\theta d\phi^2)), \quad k = 0 \\
ds^2 & = & -dt^2 + a^2(t)(d\chi^2 + \sinh^2\chi(d\theta^2 + \sin^2\theta d\phi^2)), \quad k = -1
\end{eqnarray}
where $t$ is the cosmic time since the Big Bang, $a(t)$ is the expansion
parameter, and individual galaxies participating in the cosmic expansion
follow geodesics with constant values of $\chi$, $\theta$, and $\phi$. 
These three are called co-moving coordinates.  Neglecting peculiar
velocities, galaxies remain at constant positions in co-moving coordinates
as the universe expands.  Now $a(t)$ obeys Friedmann's equations:
\begin{eqnarray}
(\frac{a_{,t}}{a})^2 & = & -\frac{k}{a^2} + \frac{\Lambda}{3} + \frac{8\pi\rho_m}{3} + \frac{8\pi\rho_r}{3} \\
2(\frac{a_{,tt}}{a}) & = & \frac{2\Lambda}{3} - \frac{8\pi\rho_m}{3} - \frac{16\pi\rho_r}{3}
\end{eqnarray}
where $\Lambda = const.$, is the cosmological constant, $\rho_m \propto
a^{-3}$, is the average matter density in the universe, including cold dark
matter, $\rho_r \propto a^{-4}$ is the average radiation density in the
universe, primarily the cosmic microwave background radiation.  The second
equation shows that the cosmological constant produces an acceleration in
the expansion while the matter and radiation produce a deceleration.  Per
unit mass density, radiation produces twice the deceleration of normal
matter because positive pressure is gravitationally attractive in Einstein's
theory and radiation has a pressure in each of the three directions
$(x,y,z)$ which is $1/3^\mathrm{rd}$ the energy density. 

We can define a conformal time $\eta$ by the relation $d\eta = dt/a$, so that
\begin{equation}
\eta(t) = \int_0^t \frac{dt}{a}
\end{equation}
Light travels on radial geodesics with $d\eta = \pm d\chi$ so a
galaxy at a co-moving distance $\chi$ from us emitted the light we see today
at a conformal time $\eta(t) = \eta(t_0) - \chi$.  Thus, we can calculate the
time $t$ and redshift $z = a(t_0)/a(t) - 1$ at which that light was emitted. 
Conversely, if we know the redshift, given a cosmological model (i.e. values
of $H_0$, $\Lambda$, $\rho_m$, $\rho_r$, and $k$ today) we can calculate the
co-moving radial distance of the galaxy from us from its redshift, again
ignoring peculiar velocities. For a more detailed discussion of distance
measures in cosmology, see \cite{hogg99}.

The WMAP satellite has measured the cosmic microwave background in exquisite
detail (\cite{wmap}) and combined this data with other data
(\cite{x2dFpower,x2dFbias,lyalpha1,lyalpha2,gar98,sn1,hst,sn2})
to produce accurate data on the cosmological model (\cite{wmapcosmo}). 
We adopt best fit values at the present epoch, $t = t_0$, based on the WMAP data of:
\begin{eqnarray}
H_0            & \equiv & \frac{a_{,t}}{a} = 71 \,\mathrm{\frac{km}{sec\ Mpc}}, \nonumber \\
\Omega_\Lambda & \equiv & \frac{\Lambda}{3H_0^2} = 0.73, \nonumber \\
\Omega_r       & =      & 8.35 \cdot 10^{-5}, \nonumber \\
\Omega_m       & \equiv & \frac{8\pi \rho_m}{3 H_0^2} = 0.27 - \Omega_r, \nonumber \\
k & = & 0. \nonumber
\end{eqnarray}
The WMAP data implies that
$w \approx -1$ for dark energy (ie. $p_{vac} = w \rho_{vac} \approx
-\rho_{vac}$), suggesting that a cosmological constant is an excellent model for
the dark energy, so we are simply adopting that.  The current Hubble radius
$R_{H_o} = c H_o^{-1} = 4220 \,\mathrm{Mpc}$.  The cosmic microwave
background is at a redshift $z = 1089$.  Substituting, using geometrized units
in which $c = 1$, and integrating the first Friedmann equation we find the
conformal time may be calculated:
\begin{equation}
\eta(t) = \int_0^t \frac{dt}{a} = 
	\int_0^{a(t)} \{-k a^2 + \frac{8\pi}{3}a^4[\rho_m(a) + \rho_r(a)] + \frac{\Lambda}{3}a^4\}^{-1/2} da
\end{equation}
where $\rho_m \propto a^{-3}$ and $\rho_r \propto a^{-4}$. This formula will
accurately track the value of $\eta(t)$, providing that this is interpreted as
the value of the conformal time since the end of the inflationary period at
the beginning of the universe.  (During the inflationary period at the
beginning of the universe, the cosmological constant assumed a large value,
different from that observed today, and the formula would have to be changed
accordingly.  So we simply start the clock at the end of the inflationary
period where the energy density in the false vacuum [large cosmological
constant] is dumped in the form of matter and radiation.  Thus, when we
trace back to the big bang, we are really tracing back to the end of the
inflationary period.  After that, the model does behave just like a standard
hot-Friedmann big bang model.  This standard model might be properly
referred to as an inflationary-big bang model, with the inflationary epoch
producing the Big Bang explosion at the start.) Now, $a(t)$ is the radius of
curvature of the universe for the $k = +1$ and $k = -1$ cases, but for the $k = 0$
case, which we will be investigating first and primarily, there is no scale
and so we are free to normalize, setting $a(t_0) = R_{H_0} = c H_0^{-1} = 4220 \,\mathrm{Mpc}$.
Then, $\chi$ measures co-moving distances at the present epoch in units of the
current Hubble radius $R_{H_o}$.  Thus, for the $k = 0$ case, using geometrized
units, we have:
\begin{equation}
\eta(a) = \eta(a(t)) = \int_0^a (\frac{a}{a_0}\Omega_m + \Omega_r + (\frac{a}{a_0})^4\Omega_\Lambda)^{-1/2} \frac{da}{a_0}
\end{equation}
where $\Omega_m$, $\Omega_\Lambda$, $\Omega_r$ are the values at the current epoch.
Given the values adopted from WMAP we find: 
\begin{equation}
	\eta(a_0) = 3.38
\end{equation}
That means that when we look out now at $t = t_0$ (when $a = a_0$) we can see out
to a distance of
\begin{equation}
	\chi = 3.38
\end{equation}
or a co-moving distance of 
\begin{equation}
	\chi R_{H_0} = 3.38 R_{H_0} = 14,300 \,\mathrm{Mpc}.
\end{equation}
This is the effective particle horizon, where we are seeing particles at the
moment of the Big Bang.  This is a larger radius than 13.7 billion light
years -- the age of the universe (the lookback time) times the speed of light --
because it shows the co-moving distance the most distant particles we can
observe now will have from us when they are as old as we are now, i.e.
measured at the current cosmological epoch.  We may calculate the value of $\eta$
as a function of $a$, or equivalently as a function of observed redshift $z =
(a_0/a) - 1$.  Recombination occurs at $z_{rec} = 1089$, which is the redshift of
the cosmic microwave background seen by WMAP.
\begin{equation}
	\eta(z_{rec}) =  0.0671
\end{equation}
So, the co-moving radius of the cosmic microwave background is: 
\begin{equation}
	\chi R_{H_0} = (\eta_0 - \eta_{rec}) R_{H_0} = 14,000 \,\mathrm{Mpc}
\end{equation}
That is the radius at the current epoch, so at recombination the WMAP sphere
has a physical radius that is 1090 times smaller or about $13$ Mpc.

According to SDSS luminosity function data (Michael Blanton, private
communication), $L_*$ in the Press and Schechter luminosity function in B
band is $7.1 \cdot 10^9 L_{\odot}$ for $H_0 = 71
\mathrm{\,km/s\, Mpc}^{-1}$ and the mean separation between galaxies brighter
than $L_*$ is $4.1$\,Mpc. The Milky Way has $ 9.4
\cdot 10^9 L_{\odot}$ in B. Since the radius of the observable universe (out
to the cosmic microwave background) is 14 Gpc, that means that the number of
bright galaxies (more luminous than $L_*$) forming within the currently
observable universe is of order 170 billion. If our galaxy has of order 200
billion stars, the mean blue stellar luminosity is of order 0.05 $L_{\odot}$
and the mean number density of stars is at least of order $2.6 \cdot
10^9 $Mpc$^{-3}$.
Ellipticals and S0 galaxies have a higher number of stars per Solar
luminosity than the Milky Way, so a conservative estimate for the mean
number density of stars might be $5 \cdot
10^9\,\mathrm{stars/Mpc}^{3}$. Thus, the currently observable universe is
home to of order $6 \cdot 10^{22}$ stars.

\begin{deluxetable}{rrl}
\tablecaption{Co-moving radii for different redshifts\label{tblredshifts}}
\tablehead{
	\colhead{$z$} & \colhead{$r(z)$ (Mpc)} & \colhead{Note}
}
\startdata
$\infty$ & 14,283 & Big Bang (end of inflationary period)  \\
$3233$   & 14,165 & Equal matter and radiation density epoch  \\
$1089$   & 14,000 & Recombination  \\
$6$      &  8,422 & \\
$5$      &  7,933 & \\
$4$      &  7,305 & \\
$3$      &  6,461 & \\
$2$      &  5,245 & \\
$1$      &  3,317 & \\
$0.5$    &  1,882 & \\ 
$0.2$    &    809 & \\ 
$0.1$    &    413 & 
\enddata
\end{deluxetable}

We may compute co-moving radii $r = \chi R_{H_0}$ for different redshifts,
as shown in table~\ref{tblredshifts}. We can also calculate the value of
$\eta(t = \infty) = 4.50$ which shows how far a photon can travel in
co-moving coordinates from the inflationary Big Bang to the infinite future. 
Thus, if we wait until the infinite future we will eventually be able to see
out to a co-moving distance of
\begin{equation}
	r_{t=\infty} = 4.50 R_{H_0} = 19,000 \,\mathrm{Mpc}
\end{equation}
This is the co-moving future visibility limit.  No matter how long we wait,
we will not be able to see further than this.  This is surprisingly close. 
The number of galaxies we will eventually ever be able to see is only larger
than number observable today by a factor of $(r_{t=\infty}/r_{t_0})^3 = 2.36$.

This calculation assumes the false vacuum state (cosmological constant)
visible today remains unchanged.  (The WMAP data is consistent with a value
of $w = -1$, indicating that the vacuum state (dark energy) today is well
approximated by a positive cosmological constant.  This false vacuum state
(with $p_{vac} = w\rho_{vac} = -1 \rho_{vac}$) may decay by forming bubbles of normal zero density vacuum
($\Lambda = 0$) or even decay by forming bubbles of negative energy density vacuum
($\Lambda < 0$). If the present false vacuum is only metastable it will eventually
decay by the formation of bubbles of normal or negative energy density
vacuum and eventually one of these bubbles will engulf the co-moving
location of our galaxy. But if these bubbles occur late ($>10^{100}$ yrs) as
expected they will make a negligible correction to how far away in comoving
coordinates we will eventually be able to see. For a fuller discussion, see
Gott, Juri\'c et al. (astro-ph/0310571v1) and references therein.

\cite{lin90} and \cite{gv98}
 have pointed out that if the current vacuum state is the lowest
stable equilibrium then quantum fluctuations can form bubbles of high
density vacuum that will start a new inflationary epoch, new baby universes
growing like branches off a tree.  Still, as in the above case, we expect to
be engulfed by such a new inflating region only at late times (say at least $10^{100}$ years
from now) and the observer will still be surrounded by an event horizon with
a limit of future visibility in co-moving coordinates in our universe that
is virtually identical with what we have plotted.  Thus, although the future
history of the universe will be determined by the subsequent evolution of
the quantum vacuum state (as also noted by \cite{ks00}), in
practice we expect the current vacuum to stay as is for considerably longer than the
Hubble time, and in many scenarios this leaves us with a limit of future
visibility that is for all practical purposes just what we have plotted. 

If we send out a light signal now, by $t = \infty$ it will reach a radius 
$\chi = \eta(t = \infty) - \eta(t_0) = 4.50 - 3.38 = 1.12$, or
\begin{equation}
	r = 4,740 \,\mathrm{Mpc}
\end{equation}
to which we refer to as the ``outward limit of reachability''. We cannot
reach (with light signals or rockets) any galaxies that are further away
than this (\cite{bus03}).  What redshift does this correspond to?  Galaxies we observe
today with a redshift of $z = 1.69$ are at this co-moving distance. 
Galaxies with redshifts larger than $1.69$ today are unreachable. This is
a surprisingly small redshift.

We can see many galaxies at redshifts larger than $1.69$ that we will never
be able to visit or signal.  In the accelerating universe, these galaxies
are accelerating away from us so fast that we can never catch them. The
total number of stars that our radio signals will ever pass is of order $2 \times
10^{21}$.

\section{A Map Projection for the Universe}

We will choose a conformal map that will cover the wide range of scales from
the Earth's neighborhood to the cosmic microwave background.  First we will
consider the flat case ($k = 0$) which the WMAP data tells us is the
appropriate cosmological model. Our map will be two dimensional so that it
can be shown on a wall chart. \cite{lgh86} showed with their
slice of the universe, just how successful a slice of the universe can be in
illustrating large scale structure. The Sloan Digital Survey should
eventually include spectra and accurate positions for about 1 million galaxies and quasars in a 3D sample
(see \cite{sdss:edr,sdss:dr1,sdss:galaxies,sdss:quasars,sdss:lrg} for SDSS scientific
results and \cite{sdss:tiling,sdss:photmonitor,sdss:photsys,sdss:astrometry} for further
technical reference).  But
virtually complete already is an equatorial slice $4$ degrees wide 
($-2\degrees < \delta < 2\degrees$) centered on the celestial equator covering both northern and southern
galactic hemispheres.  This shows many interesting features including many
prominent voids and a great wall longer than the great wall found by \cite{gh89}.

\begin{figure}
\plotone{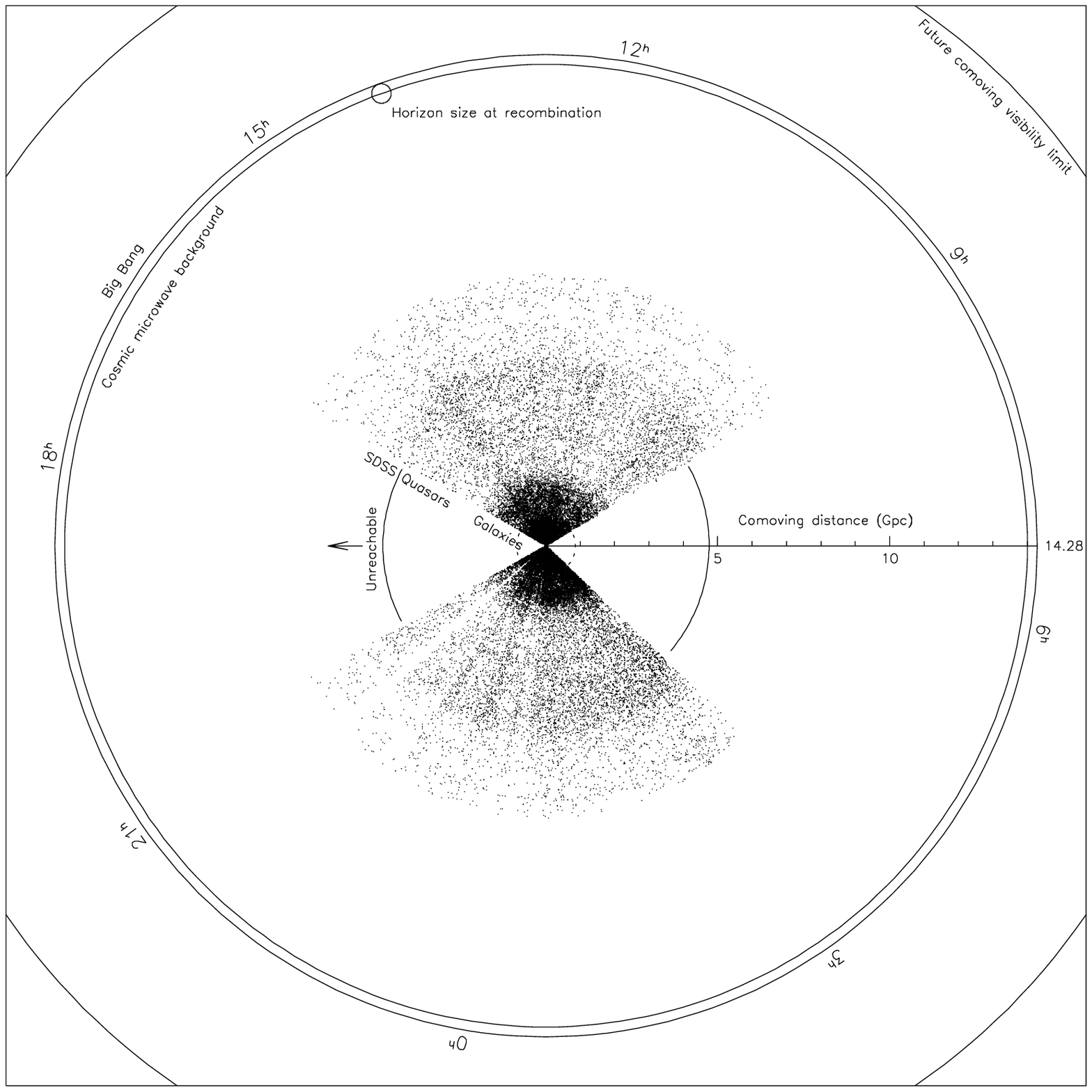}
\caption{
	Galaxies and quasars in the equatorial slice ($-2\degrees < \delta <
2\degrees$) of the Sloan Digital Sky Survey displayed in co-moving
coordinates out to the horizon. The co-moving distances to galaxies are
calculated from measured redshift, assuming Hubble flow and WMAP cosmological
parameters. This is a conformal map -- it preserves shapes. While this
map can conformally show the complete Sloan survey, the majority of
interesting large scale structure is crammed into a blob in the center.
The dashed circle marks the outer limit of figure~\ref{cpolsmall}. The
circle labeled 'Unreachable' marks the distance beyond which we cannot reach
(i.e. we cannot reach with light signals any object that is further away). This radius
corresponds to a redshift of $z = 1.69$. As 'Future comoving visibility
limit' we label the co-moving distance to which a photon would travel from
the inflationary Big Bang to the infinite future. This is the maximum radius
out to which observations will ever be possible. At $4.50 R_{H_0}$, it is
suprisingly close.
\label{cpolbig}}
\end{figure}

Since the observed slice is already in a flat plane ($k = 0$ model, along the
celestial equator) we may project this slice directly onto a flat sheet of
paper using polar coordinates with $r = \chi R_{H_0}$ being the co-moving distance, and $
\theta$ being the right ascension.  (CMB observations from Boomerang, DASI, MAXIMA and WMAP
indicate that the case $k = 0$ is the appropriate one for the universe.  For
mathematical completeness we will also consider the $k = +1$ and $k = -1$ cases
in an appendix.) We wish to show large scale structure and the extent of the
observable universe out to the cosmic microwave background radiation
including all the SDSS galaxies and quasars in the equatorial slice.  In
figure~\ref{cpolbig}, one can see the cosmic microwave background at the surface of last
scattering as a circle.  Its co-moving radius is $14.0$~Gpc.  (Since the
size of the universe at the epoch of recombination is smaller that that a
present by a factor of $1 + z = 1090$, the true radius of this circle is
about $12.84$~Mpc.) Slightly beyond the cosmic microwave background in
co-moving coordinates is the Big Bang at a co-moving distance of $14.3$~Gpc.

(Imagine a point on the cosmic microwave background circle.  Draw a radius
around that point that is tangent to the outer circle labeled Big Bang, as
shown in the figure, in other words, a circle that has a radius equal to the
difference in radius between the cosmic microwave background circle and the
Big Bang circle.  That circle has a co-moving radius of $283$~Mpc. That is
the co-moving horizon radius at recombination. If the Big Bang model --
without inflation -- were correct we would expect a point on the cosmic
microwave background circle to be causally influenced only by things inside
that horizon radius at recombination.  The angular radius of this small
circle as seen from the Earth is $(283 \,\mathrm{Mpc} / 14,000 \,\mathrm{Mpc})$
radians or $1.16\degrees$. If the Big Bang model without inflation were correct
we would expect the cosmic microwave background to be correlated on scales
of at most $1.16\degrees$.  Inflation, by having a short period of accelerated
expansion during the first $10^{-34}$ seconds of the universe, puts distant
regions in causal contact because of the slight additional time allowed when
the universe was very small.  So, with inflation, we can understand
why the cosmic microwave background is uniform to one part in 100,000 all
over the sky.  Furthermore, random quantum fluctuations predicted by
inflation add a series of adiabatic fluctuations which are expected to have
a peak in the power spectrum at an angular scale about the size of the
horizon radius at recombination calculated above, $\sim 0.86$ degrees.)

Beyond the Big Bang circle is the circle showing the future co-moving
visibility limit.  If we wait until the infinite future, we will be able to
see out to this circle.  (In other words, in the infinite future, we will be
able to see particles at the future co-moving visibility limit as they
appeared at the Big Bang.)

The SDSS quasars extend out about halfway out to the cosmic microwave
background radiation.  The distribution of quasars shows several features. 
The radial distribution shows several shelves due to selection effects as
different spectral features used to identify quasars come into view in the
visible.  Several radial spokes appear due to incompleteness in some narrow
right ascension intervals.  Two large fan shaped regions are empty and not
surveyed because they cover the zone of avoidance close to the galactic
plane which is not included in the Sloan survey.  These excluded regions run from approximately
$3.7\,\mathrm{h} \lesssim \alpha \lesssim 8.7 \,\mathrm{h}$
and approximately $16.7\,\mathrm{h} \lesssim \alpha \lesssim 20.7\,\mathrm{h}$.  The
quasars do not show noticeable clustering or large scale structure. This is
because the quasars are so widely spaced that the mean
distance between quasars is larger than the correlation length at that
epoch.

The circle of reachability is also shown. Quasars beyond this circle are
unreachable. Radio signals emitted by us now will only reach out as far as
this circle, even in the infinite future.

\begin{figure}
\plotone{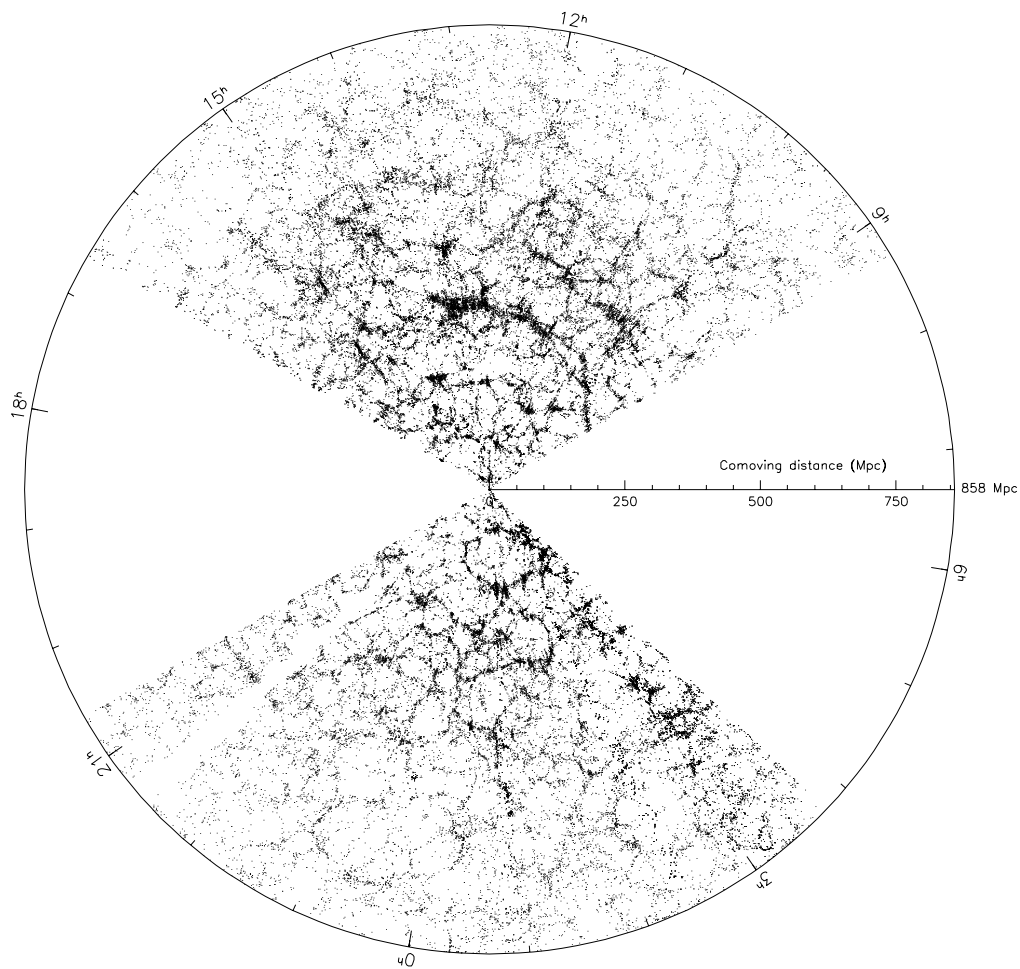}
\caption{
	Zoom in of the region marked by the dashed circle in figure
\ref{cpolbig}, out to 0.06 $r_{\mathrm{horizon}}$ ($= 858$~Mpc). 
The points shown are galaxies from the main and bright red galaxy samples of
the SDSS. Compared to figure \ref{cpolbig}, we can now see a lot of
interesting structure. The Sloan Great Wall can be seen stretching from
$8.7$h to $14$h in R.A. at a median distance of about $310$~Mpc. Although
the large scale structure is easier to see, a "zoom in" like this fails to
capture and display, in one map, the sizes of modern redshift surveys.
\label{cpolsmall}
}
\end{figure}

The SDSS galaxies appear as a black blob in the center.  There is much
interesting large scale structure here but the field is too crowded and
small to show it.  This illustrates the problem of scale in depicting the
universe.  If we want a map of the entire observable universe on one page,
at a nice scale, the galaxies are crammed into a blob in the center.  Let us
enlarge the central circle of radius $0.06$ times the distance to the Big Bang
circle by a factor of 16.6 and plot it again in figure~\ref{cpolsmall}.  This now shows a
circle of co-moving radius $858$~Mpc.  Almost all of these points are galaxies
from the galaxy and bright red galaxy samples of the SDSS.  Now we can see a
lot of interesting structure.  The most prominent feature is a Sloan Great Wall at
a median distance of about $310$~Mpc stretching from $8.7$h to $14$h in R.A.  There are
numerous voids. A particularly interesting one is close
in at a co-moving distance of 125 Mpc at 1.5h R.A. At the far end of this
void are a couple of prominent clusters of galaxies which are recognizable
as "fingers of God" pointing at the Earth. Redshift in this map is taken as
the co-moving distance indicator assuming participation in the Hubble flow,
but galaxies also have peculiar velocities and in a dense cluster with a
high velocity dispersion this causes the distance errors due to these
peculiar velocities to spread the galaxy positions out in the radial
direction producing the "finger of God" pointing at the Earth.  Numerous
other clusters can be similarly identified.  This is a conformal map, that
preserves shapes -- excluding the small effects of peculiar velocities.  The
original CfA survey in which Geller and Huchra discovered the Great Wall had
a co-moving radius of only 211 Mpc, which is less than a quarter of the
radius shown in figure~\ref{cpolsmall}.  Figure~\ref{cpolsmall} is a quite impressive
picture, but it does not capture all of the Sloan Survey.  If we displayed
figure~\ref{cpolbig} at a scale enlarged by a factor of 16.6 the central portion of the
map would be as you see displayed at the scale shown in figure 2 which is
adequate, but the Big Bang circle would have a diameter of 6.75~feet.  You
could put this on your wall, but if we were to print it in the journal for
you to assemble it would require the next 256 pages.  This
points out the problem of scale for even showing the Sloan Survey all on one
page.  Small scales are also not represented well.  The distance to the
Virgo Cluster in figure~\ref{cpolsmall} is only about 2~mm and the distance from the Milky
Way to M31 is only $1/13$th of a millimeter and therefore invisible on this
Map. Figure~\ref{cpolsmall}, dramatic as it is, fails to capture a picture of all the
external galaxies and quasars.  The nearby galaxies are too close to see and
the quasars are beyond the limits of the page.

\begin{figure}
\plotone{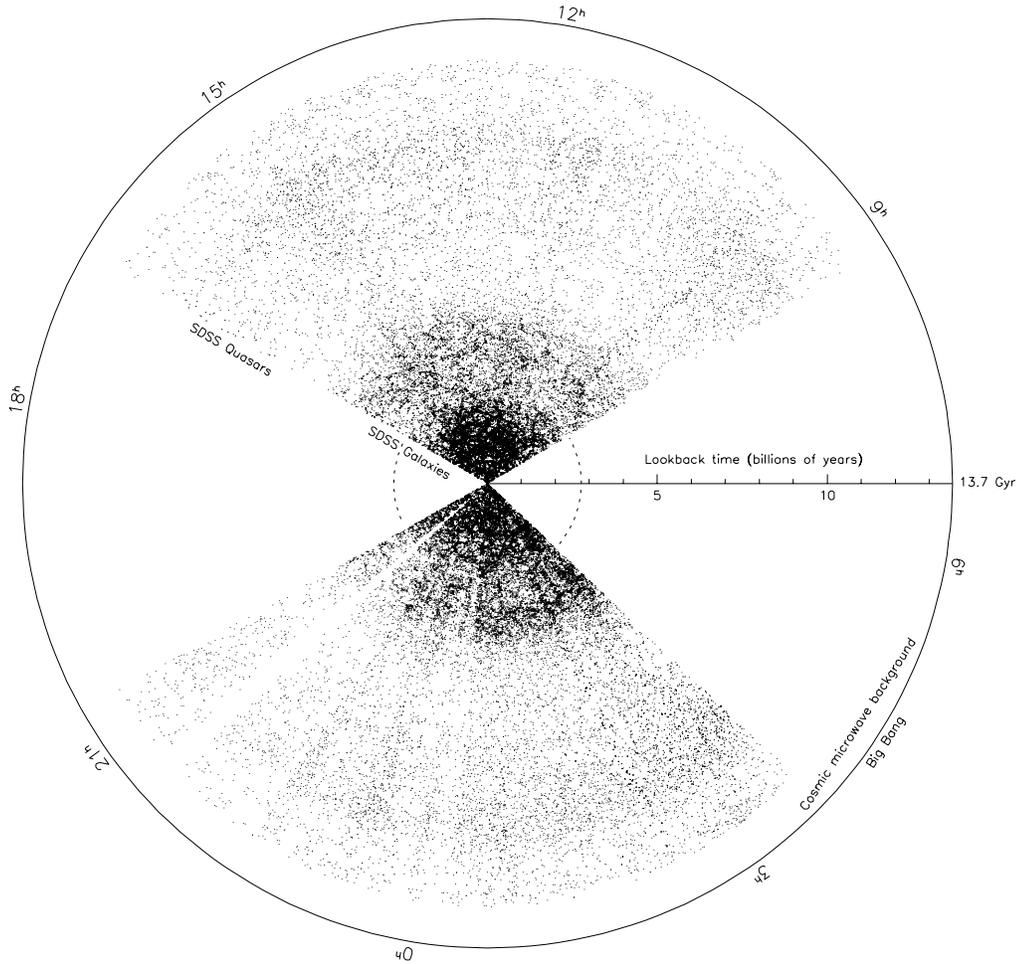}
\caption{
	Galaxies and quasars in the equatorial slice of the SDSS, displayed
in lookback time coordinates. The radial distance in the figure corresponds
to lookback time. While the Galaxies at the center occupy a larger area,
 this map is a misleading portrayal as far as shapes and the geometry of
space are concerned. It is not conformal -- it compresses the area close to
the horizon (this compression is more explicitly shown in
figure~\ref{grid}). Also, the galaxies are still too crowded in the center
of the map to show all of the intricate details of their clustering.
Figure~\ref{polsmall} shows a zoom in of the region inside the dashed
circle.
\label{polbig}
}
\end{figure}

\begin{figure}
\plotone{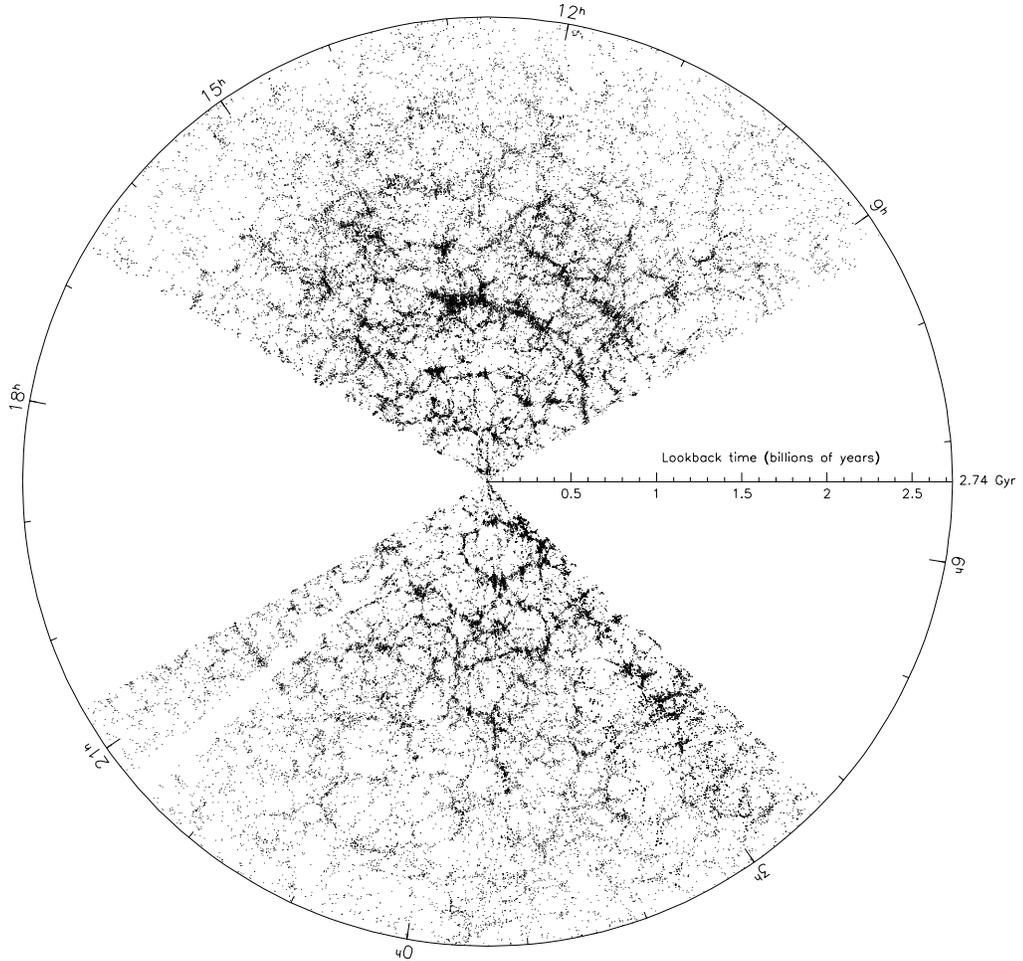}
\caption{
	Zoom in of the region marked by the dotted circle in figure
\ref{polbig}, showing SDSS galaxies out to 0.2 $t_{\mathrm{horizon}}$.
The details of galaxy clustering are now displayed much better. However,
like figure~\ref{cpolsmall}, it still fails to capture the whole survey
in one, reasonably sized, map.
\label{polsmall}
}
\end{figure}

\begin{figure}
\plotone{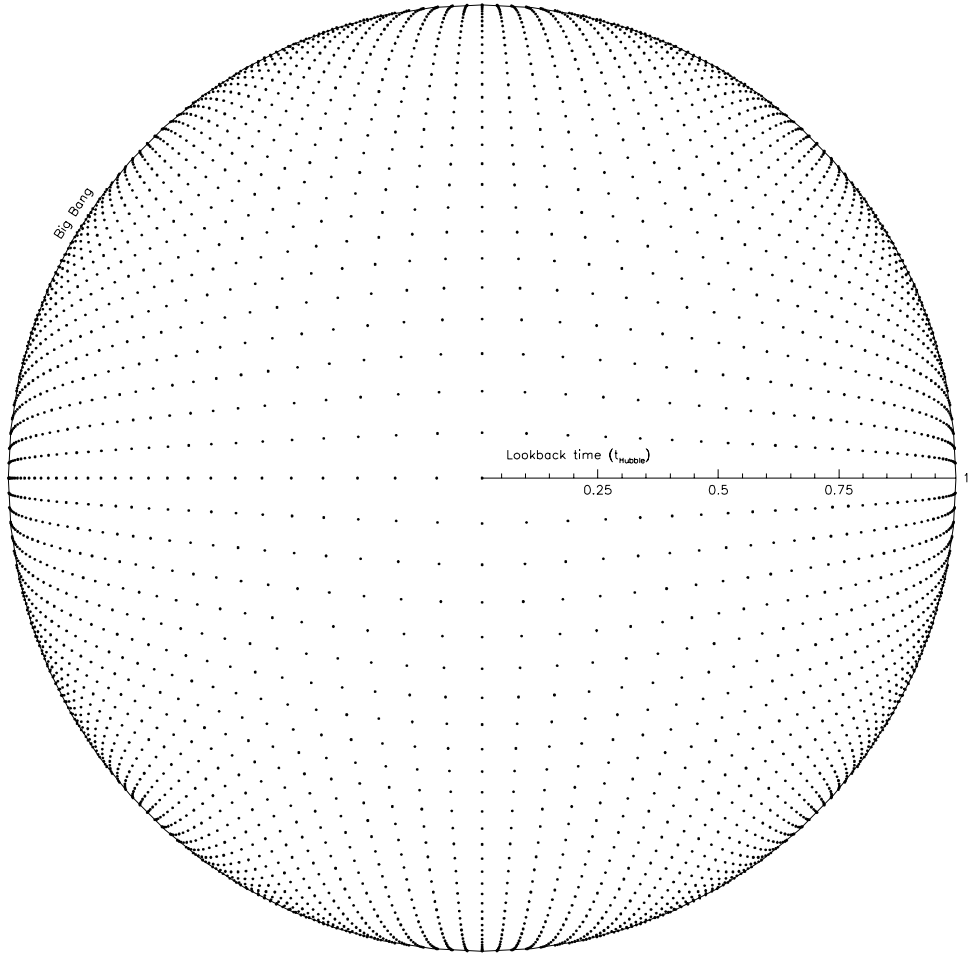}
\caption{
	Square comoving grid shown in lookback time coordinates.
	Grid spacing is $0.1 R_{H_0} = 422.24$~Mpc. 
	Each grid square would contain an equal number of
galaxies in a flat slice of constant vertical thickness.
The distortion of space that is produced by using the lookback time
is obvious as the squares become more and more distorted in
shape as one approaches the horizon.
\label{grid}
}
\end{figure}

We may try plotting the universe in lookback time rather than co-moving
coordinates.  The result is in figure~\ref{polbig}.  The outer circle is the
cosmic microwave background.  It is indistinguishable from the Big Bang as
the two are separated by only $380,000$ years out of $13.7$ billion years. 
The SDSS quasars now extend back nearly to the cosmic microwave background
radiation (since it is true that we are seeing back to within a billion
years of the Big Bang).  Lookback time is easier to explain to a lay
audience than co-moving coordinates and it makes the SDSS data look more
impressive, but it is a misleading portrayal as far as shapes and the
geometry of space are concerned.  It misleads us as to how far out we are
seeing in space.  For that, co-moving coordinates are appropriate. figure~\ref{polbig} 
does not preserve shapes -- it compresses the large area
between the SDSS quasars and the cosmic microwave background into a thin
rim. This is not a conformal map.  The SDSS galaxies now occupy a larger
space in the center, but they are still so crowded together that one can
not see the large scale structure clearly.  Figure~\ref{polsmall} shows the
central 0.2 radius circle (shown as a dotted circle in figure~\ref{polbig})
enlarged by a factor of 5.  Thus if we were to make a wall map of the
observable universe using lookback time at the scale of
figure~\ref{polsmall} it would only need to be 2 feet across and would only
require the next 25 pages in the journal to plot.  This is an advantage of
the lookback time map.  It makes the interesting large scale structure that
we see locally (figure~\ref{polsmall}) a factor of slightly over 3 larger in
size relative to the cosmic microwave background circle than if we had used
co-moving coordinates. Figure~\ref{polsmall} looks quite similar to
figure~\ref{cpolsmall}.  At co-moving radii less than 858 Mpc, the lookback
time and co-moving radius are rather similar.  Still, figure~\ref{polsmall}
is not perfectly conformal. Near the outer edges there is a slight radial
compression that is beginning to occur in the lookback time map as one
goes toward the Big Bang.  The effects of radial compression are illustrated in
figure~\ref{grid}, where we have plotted a square grid in co-moving
coordinates in terms of lookback time as would be depicted in
figure~\ref{polbig}.  Each grid square would contain an equal number of
galaxies in a flat slice of constant vertical thickness.  This shows the
distortion of space that is produced by using the lookback time. The squares become more and more distorted in
shape as one approaches the edge.

Thus, it would be useful to have a conformal map projection that
would show the whole SDSS survey, including galaxies, quasars and the cosmic
microwave background, as well as smaller scales, covering the local
supercluster, the Local Group, the Milky Way, nearby stars, the Sun and
planets, the Moon and artificial Earth satellites.  Such a map is possible.

Consider the complex plane $(u,v) = u + \imag v$ where $\imag = \sqrt{-1}$ and $u$ and $v$ are real
numbers.  Every complex number $W = u + \imag v$ will be represented as a point in
the $(u,v)$ plane where $u$ and $v$ are the usual Cartesian coordinates.  The
function $Z = \imag \ln(W)$ maps the plane $(u,v)$ onto the plane $(x,y)$ where $Z = x + \imag y$. 
The $(u,v)$ plane represents a slice of the universe in isotropic
coordinates (in this case co-moving coordinates since $k = 0$), and the $(x,y)$
plane represents our map of the universe.  The inverse function $W = \exp(Z/\imag)$
is the inverse map that takes a point in our map plane $(x, y)$ back to the
point it represents in the universe $(u, v)$.  In the universe it is useful to
establish polar coordinates $(r, \theta)$ where
\begin{eqnarray}
u & = & r \cos \theta \\
v & = & r \sin \theta
\end{eqnarray}
and $r = (u^2 + v^2)^{1/2}$ is the (co-moving) distance from the center of the
Earth and $\theta = \arctan(v/u)$ is the right ascension measured in radians.  Since
\begin{equation}
\cos \theta = \frac{\e^{\imag\theta} + \e^{-\imag\theta}}{2}, \quad
\sin \theta = \frac{\e^{\imag\theta} - \e^{-\imag\theta}}{2\imag}
\end{equation}
it is clear that
\begin{eqnarray}
W = u + \imag v = r(\cos \theta + \imag \sin\theta) & = & r \e^{\imag\theta}          \\
Z = \imag \ln(W) = i (\ln r + \imag\theta) = -\theta + \imag \ln r & = & x + \imag y
\end{eqnarray}
so:
\begin{eqnarray}
x & = & -\theta \\
y & = & \ln r
\end{eqnarray}
Thus, the entire $(u,v)$ plane, except the origin $(0,0)$, is mapped into an
infinite vertical strip of horizontal width $2\pi$, i.e.
\begin{equation}
	-2\pi < x \le 0, \quad  -\infty < y < \infty
\end{equation}

\begin{figure}
\plotone{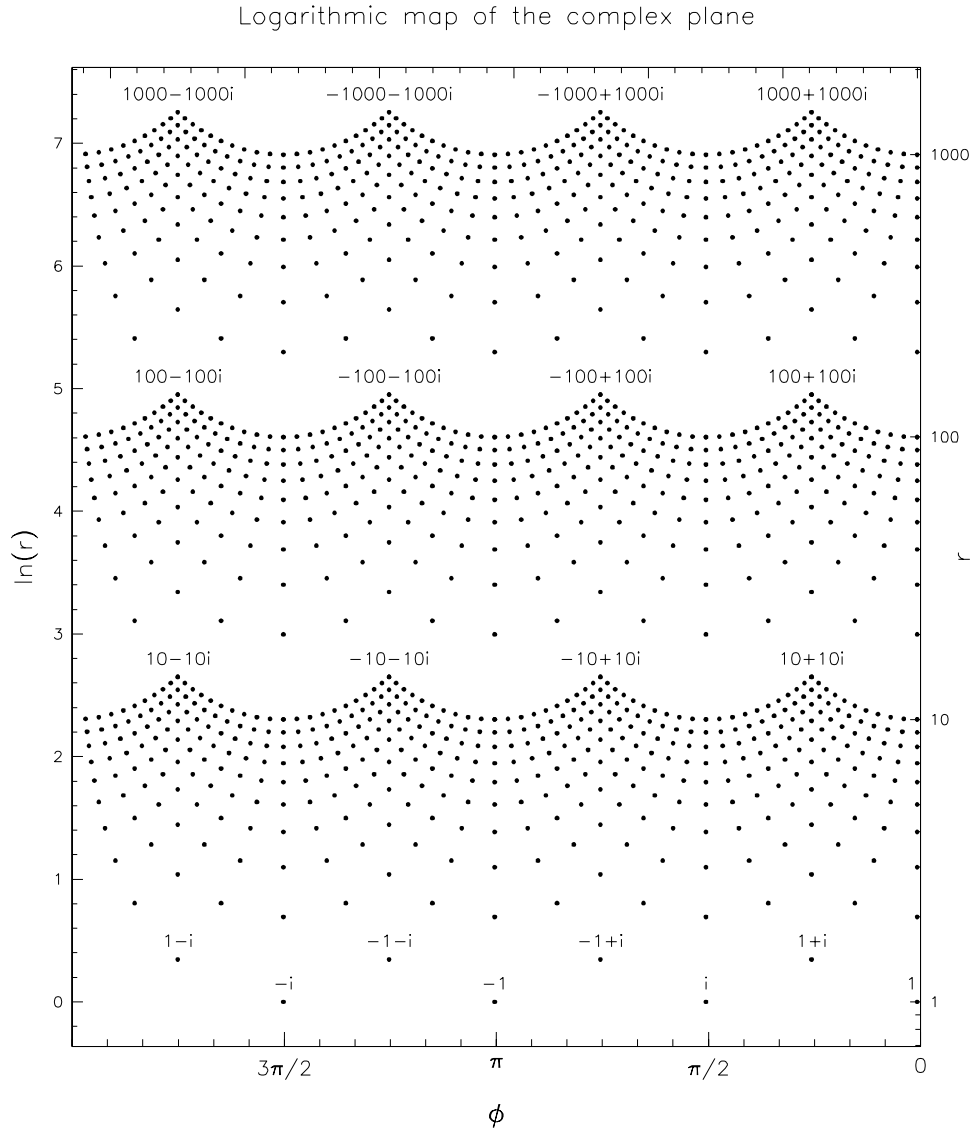}
\caption{
	Logarithmic map of the complex plane. Vertical axis is the natural
logarithm of the absolute value of a complex number, while its phase is
plotted on the horizontal axis. We plot $4 \times 10 \times 10$ numbers per
decade, from the first three decades. This map is conformal, but covers a wide range of scales.
\label{complex}
}
\end{figure}

(Figure~\ref{complex} shows the complex plane $u + \imag v$ mapped onto the $x + \imag y$ plane by this
map.  One can take this map and make it into a slide rule for multiplying
complex numbers.  Photocopy the map on this page and cut it out.  Tape the
left hand edge to the right hand edge to make a paper cylinder.  The $\theta$
coordinate now measures longitude angle on that cylinder.  Now photocopy the
map onto a transparency, and cut it out, and again tape the left hand edge
to the right hand edge to make a transparent cylinder.  In cutting out the
left and right hand sides of the map cheat a little, cut along the outside
edges of the map borderlines so that the circumference of the transparency
cylinder is just a tiny bit larger than the paper cylinder and so that it
will fit snugly over it.  With the paper cylinder snugly inside the
transparent cylinder you are ready to multiply.  If you want to multiply two
numbers $A = a + b\imag$ and $C = c + d\imag$ all you do is rotate and slide the
transparent cylinder until the transparency number 1 (i.e. $1 + 0\imag$) is directly over
the number $a + b\imag$ on the paper cylinder, then look up the number $c + d\imag$ on
the transparent cylinder, below it on the paper cylinder will be the product
$A \cdot B$.  The logarithm of $AB$ is equal to the sum of the logarithms of $A$ and $B$. 
Of course, on the real axis, $\theta = 0$ ($x = 0$), the map looks like the scale on
a slide rule. Alternatively make a flat slide rule for multiplying complex
numbers: make two photocopies of the map on white paper and tape them
together to make two cycles in $\theta$ from right to left.  Then make one
photocopy of the map onto a transparency.  Lay the number $1$ (on the
transparency) on top of the number $A$ in the right hand cycle of the paper 
map and look on the transparency for the number $B$, below it on the paper 
map will be the product $AB$.)

\begin{figure}
\epsscale{1.3}
\plotone{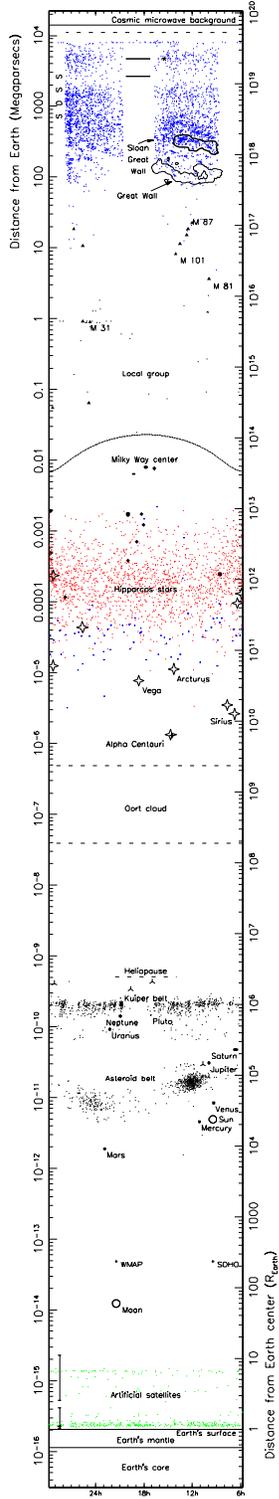}
\caption{Pocket map of the universe\label{pocketmap}. A smaller version of the
more detailed map shown in the foldout figure~\ref{p0}}
\end{figure}

For convenience on our map of the universe, let $r = (\chi R_{H_0})/r_{E}$ (co-moving
cosmological distance/radius of the Earth) be measured in units of the
Earth's equatorial radius $r_{E} = 6378$km.  Thus, circles of constant radius ($r = const.$)
from the center of the Earth are shown as horizontal lines ($y = const.$) 
 in the map, and rays of constant right ascension ($\theta = const.$) are
shown as vertical lines ($x = const.$) in the map. The surface of the Earth
(at its equator) is a circle of unit radius in the $(u,v)$ plane, and is the
line $y = 0$ in the map. The region $y < 0$ in the map represents the interior
of the Earth, so one can show the Earth's mantle and liquid and solid core.
The solid inner core has a radius about $0.19\,r_{E}$, thus, the lower edge of the
map must extend to $y = \ln(0.19) = -1.66$ to show it.  The region $y > 0$ shows
the universe beyond the Earth.  The co-moving future observability limit at
a radius of $19$~Gpc is at $9.2 \times 10^{19}$ Earth radii, and so the upper edge of the
map must extend to $y = \ln(9.2 \times 10^{19}) = 45.97$ to show it.  Thus the dimensions
of the map are $\Delta x = 2\pi$, and $\Delta y = 47.63$. The aspect ratio for the map is
$\mathrm{height/width} = 7.58$. See figure~\ref{pocketmap} for a small scale version of this that
will fit on one page.  (A square map would have dimensions $2\pi \times 2\pi$ and would
cover a scale ratio from bottom to top of $\exp(2\pi) = 535.49$. A map with an
aspect ratio $\mathrm{height/width} = 7.58$ covers a scale ratio from bottom to top
of $535.49^{7.58}$.)

At a scale of about 1 radian per inch for the angular scale, this would make
a map about 6.28~inches wide by 47.6~inches tall, which could be easily
displayed as a wall chart. We have presented the map at approximately this
scale later in this article.

This is not the first time logarithmic coordinates have been used for a map
of the universe.  The Amoco Map of Space Mysteries (\cite{amoco}) plotted the curved
surface of the Earth and above it altitude (from the surface, not distance
from the center) marked off in equal intervals labeled 1 mile, 10 miles, 100
miles, 1,000 miles, 10,000 miles, 100,000 miles, 1 million miles, 10 million
miles, and 100 million miles.  In this range Solar system objects from the
moon to Venus, Mars and the Sun are plotted properly.  But although $\alpha$~Centauri,
the Milky Way and M31 are shown beyond they are not shown at
correct scale (even logarithmically).  The Earth's surface is plotted where
an altitude of 0.1 miles should have been.  In any case, because of the
curvature of the Earth's surface in the map, even in the range between an
altitude of 10 miles and 100 million miles, the map is not conformal.  In
October 1999, National Geographic presented a map of the universe (that one
of us (JRG) participated in producing) which was a 3D view with a
spherical Earth at the center with equal width shells surrounding it like an
onion with radii of 400,000 miles, 40 million miles, 4 billion miles, 4
trillion miles, 10 light years, 1,000 light years, 100,000 light years, 10
million light years, 1 billion light years, 11-15 billion light years
(\cite{viewfromearth}).  This
map displays objects from the moon to the microwave background but is also
not conformal.

The map projection we are proposing is conformal because the derivatives of
the complex function $Z = \imag \ln W$ have no poles or zeros in the mapped region. 
If we want to see how a little area of the universe slice is mapped onto our
slice we should do a Taylor expansion: the point $W + \Delta W$ is mapped onto the
point $Z + \Delta Z = Z + (dZ/dW) \Delta W$ in the limit where $\Delta W \to 0$ providing that
$dZ/dW \ne \infty$ so the map doesn't blow up there and $dZ/dW \ne 0$ so that the
second and higher order terms in the Taylor expansion can be ignored
(providing that none of the higher derivatives $d^nZ / dW^n$ become infinite at
the point $W$).  In this case, in the limit as $\Delta W \to 0$, the Taylor series
is valid using just the first derivative term:
\begin{eqnarray}
\Delta Z      & = & \frac{dZ}{dW}\Delta W				\\
\frac{dZ}{dW} & = & \frac{d(\imag \ln W)}{dW} = \frac{\imag}{W}
\end{eqnarray}
Thus for $W \ne 0$ and finite (i.e. excluding the center of the Earth and the
point at infinity which are not mapped anyway) $dZ/dW$ is neither zero nor
infinity. The higher derivatives $(n \ge 2): d^nZ/dW^n = \imag (-1)^n n! W^{-n+1}$ are also
finite when $W$ is finite and non zero.  Thus, the point $W + \Delta W$ is mapped onto
the point $Z + \Delta Z = Z + dZ/dW \Delta W$ in the limit where $\Delta W \to 0$.
Characterize the point $W$ as
\begin{equation}
W = r_w(\cos\theta_w + \imag\sin\theta_w)
\end{equation}
Now the product of two complex numbers
\begin{eqnarray}
	A & = & r_a(\cos\theta_a + \imag\sin\theta_a) \nonumber \\
	B & = & r_b(\cos\theta_b + \imag\sin\theta_b) \nonumber
\end{eqnarray}
is
\begin{equation}
	A \cdot B = r_a r_b (\cos(\theta_a + \theta_b) + \imag\sin(\theta_a + \theta_b)),
\end{equation}
so
\begin{equation}
	\frac{1}{W} = \frac{1}{r_w} (\cos(-\theta_w) + \imag\sin(-\theta_w)),
\end{equation}
and since $\imag = \cos(\pi/2) + \imag\sin(\pi/2)$,
\begin{equation}
	\imag \frac{1}{W} = \frac{1}{r_w}(\cos(\frac{\pi}{2} - \theta_w) + \imag\sin(\frac{\pi}{2} - \theta_w))
\end{equation}
and 
\begin{eqnarray}
\Delta Z & = & r_{\Delta Z}(\cos\theta_{\Delta Z} + \imag\sin\theta_{\Delta Z}) \nonumber \\
	 & = & \frac{1}{r_w}(\cos(\frac{\pi}{2} - \theta_w) + \imag\sin(\frac{\pi}{2} - \theta_w)) \cdot
               r_{\Delta W}(\cos\theta_{\Delta W} + \imag\sin\theta_{\Delta W}) \nonumber \\
	 & = & \frac{r_{\Delta W}}{r_w}( \cos(\theta_{\Delta W} + \frac{\pi}{2} - \theta_w) +
               \imag\sin(\theta_{\Delta W} + \frac{\pi}{2} - \theta_w) )
\end{eqnarray}

Thus, the vector $\Delta W$ is rotated by an angle $\pi/2 - \theta_w$ and multiplied by a
scale factor of $1/r_w$.  Since any two vectors $\Delta W_1$ and $\Delta W_2$ at the point $W$
will be rotated by the same amount when they are put on the map the angle
between them is preserved in the map, and so the map projection is conformal.  The only
place the first derivative (and the higher derivatives) blow up (or go to
zero) is at $W = 0$ at the center of the Earth or at the point at infinity $W =
\infty$.  But the Earth's center does not appear on the map at all (it is at $y = -\infty$).
This is a set of measure zero.  Likewise, the point at infinity $W = \infty$ is
not plotted either (it is at $y = +\infty$).  So the map projection is conformal at
all points in the map. Shapes are preserved locally. 

The conformal map projection for showing the universe presented here was
developed by JRG in 1972, and he has produced various small versions of it
over the years.  These have been shown at various times, notably to the
visiting committee of the Hayden Planetarium in 1996 and to the staff of the
National Geographic Society in 1999.  Recent discoveries within a wide range
of scales from the solar system objects, to the SDSS galaxies and quasars
have prompted us to produce and publish the map in a large scale format.

Our large scale map is shown in figure~\ref{p0} (the foldout). A radial vector $\Delta W$ (pointing away
from the Earth's center) at the point $W$ points in the direction $\theta_{\Delta W} = \theta_w$.
This vector in the map is rotated by an angle $\pi/2 - \theta_w$ so that $\theta_{\Delta Z} = \theta_{\Delta W} +
\pi/2 - \theta_w = \pi/2$, so that it points in the vertical direction.  Small regions
in the universe are rotated in the map so that the radial direction, away
from the center of the Earth, is in the vertical direction in the map. 
Radial lines from the Earth's center are plotted as vertical lines.  Circles
of constant radius from the center of the Earth are horizontal lines.  The
length of the vector $\Delta W$ is multiplied by a scale factor $1/r_w$.  Thus, the
scale factor at a given point on the map can be read off as proportional to 1 over the
distance of the point from the Earth, $r_w$. (Objects that are twice as far
away are shown at $1/2$ scale and objects that are 10 times
further away are shown at $1/10$th the scale, and so forth).

Radial lines separated by an angle $\theta$ (in radians), going outward from the Earth will be
plotted as parallel vertical lines separated by a horizontal distance
proportional to $\theta$.
Thus objects of the same angular size in the sky (as seen from the
center of the Earth) will be plotted as the same size on the map.
The Sun and Moon which have the
same angular size in the sky ($0.5\degrees$) will be plotted as circles of the same
size on the map (since their cross sections are circles and shapes are
preserved locally in a conformal map projection).

The map gives us that Earthling's view of the universe that we want. 
Objects are shown at a size in the map proportional to the size they subtend
in the sky.  The Sun and Moon are equally large in the sky and so appear of
the same size in the map.  Objects that are close to us are more important
to us -- as depicted in that New Yorker cover.  Buildings on $9^\mathrm{th}$ avenue may
subtend as large an angle to our eye as the distant state of California. 
Our loved ones -- important to us -- are often only a few feet away and subtend
a large angular scale to our eyes.  A murder occurring in our neighborhood
draws more of our attention than a murder of someone halfway around the
globe.  Plotting objects at a size equal to their angular scale makes
psychological sense.  Objects are shown taking up an area on the map that is
proportional to the area they subtend in the sky (if they are approximately
spherical -- as many astronomical bodies are). The importance of the object in
the map (the fraction of the map it takes up) is proportional to the chance
we will see the object if we look out along a random line of sight.  Indeed,
if we look at the map from a constant distance, the angular size of the
objects in the map will be proportional to the angular size they subtend in
the sky.  The visual prominence of objects in the map will be proportional
to their visual prominence in the sky.

Of course this means that the Moon and Sun and other objects will be shown
at their true scale relative to their surroundings (i.e. the Moon is shown
in correct scale relative to the circumference of its orbit) which is
small because they are small in the sky.  Suppose we made an Mollweide equal
area map projection of the sky at a scale to fit on a journal page: an
ellipse with horizontal width 6.28~inches and vertical height of 3.14~inches.
Along the equatorial plane the scale is linear at 1~inch per radian. 
The diameter of the Moon or Sun is $1/2\degrees$, or $1/720$th of the $360\degrees$ length
around the equator.  Thus, on this sky map the Moon would have a diameter of
$(6.28 \,\mathrm{inches})/720 = 0.0087$~inches.  On our map the Moon would have
a similar diameter, for the scale of our map is approximately 1~radian = 1~inch.  The
Moon and Sun are rather small in the sky. With a printer resolution of 300
dots per inch the Sun and Moon would then be approximately 3 dots in
diameter. 
For easier visibility we have plotted the Sun and Moon as circles
enlarged by a factor of 18.
 Similar
enlargements of individual objects like the Sun and the Moon might appear on
sky maps appearing on one page.  Just as symbols for cities on world maps
may be larger than the cities themselves.  Still it is interesting to note
what the true sizes on the map should be since it shows how much empty space
in the universe there is.  M31 for example subtends an angle of about $2\degrees$ on
the sky and so would be about 0.035~inches on a map with 1~inch/radian
scale.  If versions of the map were
produced at larger scale as we shall discuss below, images of the Sun, Moon,
and nearby galaxies could be displayed at proper angular scale and simply
placed on the map.

The completed map is shown in the foldout (figure~\ref{p0}).

The map shows a complete sample of objects in the classes we are
illustrating in the equatorial slice ($-2\degrees < \delta < 2\degrees$) which is shown
conformally correct.  These objects are shown at the correct distances and right
ascensions.  This we supplement with additional famous objects out of the
plane which are shown at their correct distances and right ascensions.  So
this is basically an equatorial slice with supplements.

At the bottom, the map starts with an equatorial interior cross section of
the Earth.  First we see the solid inner core of the Earth with a radius of
$\sim 1200$~km.  Above this is the liquid outer core (1200~km -- 3480~km), and
above that are the lower (3480~km -- 5701~km) and upper mantle (5701~km --
6341~km).  The Earth's surface has an equatorial radius of 6378~km.  There is
a line designated Earth's surface (\& crust) which is a little thicker than
an ordinary line to properly indicate the thickness of the crust.  The
Earth's surface (\&~crust) line is shown as perfectly straight, because on
this scale the altitude variation in the Earth's surface is too small to be
visible. The scale at the Earth surface line is approximately $1/250,000,000$. 
The maximum thickness of the crust (37 km) is just barely visible on the
map at a resolution of 300 dots per inch, so we have shown the maximum crust
depth accordingly, by the width of the Earth surface (crust) line.

(If we had wished we could have extended the map downward to cover the
entire inner solid core of the Earth down to the central neutron (or proton)
in an iron atom located at the center of the Earth.  Since a neutron has a
radius of approximately $1.2 \times 10^{-13}$ cm or $1.9 \times 10^{-22}r_E$, the circumference of
this central neutron would be plotted as a straight line at $y = -50.0$.  The
outer circumference of the central iron atom (atomic radius of $1.40 \times 10^{-8}$cm, \cite{sla64})
would be plotted as a straight line at $y = -38.36$.  Thus, including
the entire inner solid core of the Earth down to the central neutron in an
iron atom at the center of the Earth would require (at 1~inch/radian
scale) about an additional 48~inches of map, approximately doubling its length. 
The central atom and its nucleus would then occupy the bottom 11.6~inches of
the map, giving a nice illustration of both the nucleus and all the electron
orbitals. But since we are primarily interested in astronomy, and the key
regions of the Earth's interior are covered in the map already, we have
stopped the map just deep enough to show the extent of the inner
solid core.)

We have shown the Earth's atmosphere above the Earth's surface.  The
ionosphere is shown which occupies an altitude range of 70~--~600~km.  Below the
ionosphere is the stratosphere which occupies an altitude range of 12~--~50~km. 
Although there was not enough space to include a label, the stratosphere on
the map simply occupies the tiny space between the lower error bar
indicating the bottom of the ionosphere and the "surface of the Earth" line. 
The troposphere (0~--~12~km) is of such small altitude that it is subsumed into
the Earth's surface line thickness. Above the ionosphere we have formally the exosphere, where the
mean free path is sufficiently long that individual atoms with escape
velocity can actually escape.  So the top of the ionosphere effectively
defines the outer extent of the Earth's atmosphere and the map shows
properly just how narrow the Earth's atmosphere is relative to the
circumference of the Earth.

Next we have shown all 8,420 artificial Earth satellites in orbit as
of Aug 12, 2003 (at the time of full Moon 2003/08/12 04:48 UT).  In fact
all objects in the map are shown as of that time.  This is the last full
Moon before the closest approach of Mars to the Earth in 2003.  The time was
chosen for its placement of the Sun, Moon and Mars.  We show all Earth
satellites (not just those 624 in the equatorial slice).  These are actual
named satellites, not just space junk.  Some famous satellites are
designated by name.  ISS is the International Space Station.  HST is the
Hubble Space Telescope.  These are both in low Earth orbit. 

Vanguard 1 is shown, the earliest launched satellite still in orbit.
The Chandra X-ray observatory is also shown.
There are two main altitude layers of low Earth orbiting
satellites and a scattering of them above that.  There is a quite visible
line of geostationary satellites at an altitude of 22,000~miles above the
Earth's surface. These geosynchronous satellites are nearly all in our equatorial
slice.  A surprise was the line of GPS (Global Positioning System)
satellites at a somewhat lower altitude.  These are all in nearly circular
orbits at identical altitudes and so also show up as a
line on the map. We had not
realized that there were so many of these satellites or that they would show
up on the map so prominently.  The inner and outer Van Allen radiation belts
are also shown.

Beyond the artificial Earth satellites and the Van Allen radiation belts
lies the Moon, marking the extent of direct human occupation of the universe. The
Moon is full on August 12, 2003.

Behind the full Moon at approximately 4 times the distance from Earth is the WMAP
satellite which has recently measured the cosmic microwave background.  It
is in a looping orbit about the L2 unstable Lagrange point on the opposite
side of the Sun from the Earth.  Therefore it is approximately behind the
full Moon. $180\degrees$ away, at the L1 Langrange point is the Solar and
Heliospheric Observatory (SOHO) satellite. From L1, SOHO has an unobstructed
and uninterrupted view of the Sun throughout the year.

To illustrate the distances from Earth to the nearest asteroids, we plot the
12 closest to Earth as of August 12, 2003 (labeled NEOs - Near Earth
Objects). Asteroid 2003 GY was closest to Earth at that time. Another two
them are particularly interesting. Asteroid 2003 YN107 is currently the only
known quasi-satellite of the Earth (\cite{con04}), and shall remain such
until the year 2006. Asteroid 2002 AA29 is on an interesting horseshoe orbit
(\cite{con02}, \cite{bg04}) that circulates at 1 AU and has close approaches
to Earth every 95 years. Such horseshoe orbits are typical of orbits that
have escaped from the Trojan L4, L5 points (\cite{bg04}). Since this
asteroid may have originated at 1 AU like the Earth and the great impactor
that formed the Moon, it might be an interesting object for a sample return
mission, as Gott and Belbruno have noted.

Mars is shown at approximately 9,000 Earth radii, or 0.4
astronomical units (as seen on the scale on the right).  Mars is near its
point of closest approach (which it achieved on August 27, 2003 when it was
at a center-to-center distance of 55,758,006 km). Further up
are Mercury, the Sun and Venus.  Venus is near
conjunction with the Sun on the opposite side of its orbit. The Sun is $180\degrees$
away from the Moon or halfway across the map horizontally, since the Moon is
full. The Sun is 1~AU away from the Earth.  At distances
from Earth of between $\sim 0.7$ to $\sim 5~AU$ are the main belt asteroids.
Here, out of the total of
218,484 asteroids in the ASTORB database, we have shown only those 14,183
which are in the equatorial plane equatorial slice ($-2\degrees < \delta < 2\degrees$).  If we
had shown them all, it would have been totally black. By just showing the
asteroids in the equatorial slice we are able to see individual dots. In addition, some
famous main belt asteroids, like Ceres, Eros, Gaspra, Vesta, Juno and Pallas are shown 
(even if off the equatorial slice) and indicated by name.
The width of the main belt of asteroids is shown in proper scale
relative to its circumference in the map. The belt is closer to the Earth in
the anti-solar direction since it is an annulus centered on the Sun
and the Earth is off center. Because the main belt asteroids lie
approximately in the ecliptic plane which is tilted at an angle of $23.5\degrees$
relative to the Earth's equatorial plane, there are two dense clusters where
the ecliptic plane cuts the Earth's equatorial plane and the density of
asteroids is highest.  One intersection is at about 12h and the other is at
24h.

Jupiter is shown in conjunction with the Sun approximately 6~AU from the
Earth. On either side of Jupiter we can
see the two swarms of Trojan asteroids trapped in the L4 and L5 stable
Lagrange equilibrium points $\pm 60\degrees$ away from Jupiter along its orbit.  From
the vantage point of Earth, 1~AU off center opposite Jupiter in its orbit,
the Trojans are a bit closer to the Earth than Jupiter and a bit closer to
Jupiter in the sky on each side than $60\degrees$.  The Ulysses spacecraft is visible
near Jupiter.  It is in an orbit far out of the Earth's equatorial plane,
but we have included it anyway.  Beyond Jupiter are Saturn, Uranus and
Neptune.

Halley's comet is shown between the orbits
of Uranus and Neptune, as of August 12, 2003. 

Next we show Pluto and the Kuiper belt objects.   We are showing all 772 of
the known Kuiper belt objects (rather than just those in the equatorial
plane).  It is surprising how many of them there are. Recently discovered
2003 VB12 (``Sedna''), Quaoar and Varuna, the largest KBOs currently known,
are also shown and labeled. Sedna is currently the second largest know
transneptunian object, after Pluto (\cite{btr04}). The band of Kuiper belt
objects is relatively narrow because of the selection effect that objects of
a given size become dimmer approximately as the fourth power of their
distance from the Earth and Sun. The band of Kuiper belt objects has
vertical density stripes, again due to angular selection effects depending
on where various surveys were conducted.  A sprinkling of Kuiper belt
objects extend all the way into the space between the orbit of Uranus and
Saturn (the ``Centaurs'', of which we're only plotting the ones in the
equatorial plane).

We show Pioneer 10, Voyager 1 and
Voyager 2 spacecrafts, headed away from the solar system.  These are on their
way to the heliopause, where the solar wind meets the interstellar medium. 
They have not reached it yet.

Almost a hundred times further away than the heliopause is the beginning of
the Oort cloud of comets which extends from about 8,000~AU to 100,000~AU. 
A comet entering the inner
solar system for the first time has a typical aphelion in this range.

Beyond the Oort Cloud are the stars.  The ten brightest stars visible in the
sky are shown with large star symbols.  The nearest star, Proxima Centauri
is shown with a small star symbol.  Proxima Centauri, an M5 star, is a
member of the $\alpha$~Centauri triple star system.  $\alpha$~Centauri A, at a
distance of a little over 1~pc (see scale on the left), and a solar type
star, is one of the ten brightest stars in the sky. The 10 nearest star
systems are also shown: $\alpha$ Centauri, Barnard's star, Wolf 359, Halande
21185, Sirius, UV and BL Ceti, Ross 154, Ross 248, $\epsilon$ Eridani and Lacaille
9352. Of these, $\epsilon$ Eridani has a confirmed planet circling it.

Stars with known confirmed planets circling them (with $M\sin i  < 10
M_{\mathrm{Jupiter}}$) are shown as dots with circles around them.  Of
these, 95 are solar type stars whose planets were discovered by radial
velocity perturbations.  Some of the more famous ones like 51~Peg, 70~Vir,
and $\epsilon$~Eri are labeled.  The star HD~209458 has a Jupiter-mass
planet which was discovered by radial velocity perturbations but was later
also observed in transit (\cite{maz00}, \cite{hen99}). The first planet
discovered by transit was OGLE-TR-56 which lies at a distance of over 1~kpc
from the Earth. Three other OGLE stars were also found to have transiting
planets -- TR-111, TR-113 and TR-132. These were all in the same field (R.A.
$\sim 10^\mathrm{h}50^\mathrm{m}$) at approximately the same distance ($\sim
1.5$kpc) and so would be plotted at positions on the map that would be
indistinguishable. The planet TrES-1 around GSC 02652-01324 was recently
discovered by transit and confirmed by radio velocity measurements
(\cite{alo04}). A planet circling the star OGLE 2003-BLG-235 was discovered
by microlensing (\cite{bon04}). PSR 1257+12 is a pulsar (neutron star) with
three terrestrial planets circling it which were discovered by radial
velocity perturbations on the pulsar revealed by accurate pulse timing
(\cite{wol92}, \cite{wol94}). This was the first star discovered to have
planets. Also, SO 0253+1652 is shown and labeled on the map. It is the
closest known brown dwarf (\cite{teg03}), at 3.82pc.


The first radio transmission of any significant power to escape beyond the
ionosphere was the TV broadcast of the opening ceremony of the 1936 Berlin
Olympics on August 1$^\mathrm{st}$, 1936, a fact noted by Carl Sagan in his
book \underline{Contact}. The wave front corresponding to this transmission
is a circle having a radius of $\sim 10^{11}$ Earth radii on August 12.
2004, and is indicated by the straight line labeled "Radio signals from
Earth have reached this far". Radio signals from Earth have passed stars
below this line.

The Hipparcos satellite has measured accurate parallax distances to 118,218
stars (\cite{esa97}).  We show only the 3,386 Hipparcos stars in the equatorial plane 
($-2\degrees < \delta < 2\degrees$).

Other interesting representative objects in the galaxy are illustrated:  the
Pleiades, the globular cluster M13, the Crab nebula (M1), the black hole
Cygnus X-1, the Orion Nebula (M42), the Dumbbell nebula and the Ring nebula,
the Eagle nebula, the Vela pulsar, and the Hulse-Taylor binary pulsar.
 At a distance of 8~kpc is the
Galactic Center which harbors a 2.6 million solar mass black hole.  The
outermost extent of the Milky Way optical disk is shown as a dotted line.

Beyond the Milky Way are plotted 52 currently known members of the Local Group of galaxies. 
We have included all of these, not just the ones in the equatorial plane. 
These are indicated by dots or triangles. M31's companions M32 and NGC205 are
shown as dots but not labeled since they are so close to M31.

M81, is the first galaxy shown beyond the Local Group and is a member of the
M81~--~M82 group.  Its distance was determined by Cepheid variables using the
HST.  Other famous galaxies labeled include M101, the
Whirlpool galaxy (M51), the Sombrero galaxy,
and M87, in the center of the Virgo cluster.  If
we showed all the M objects many would crowd together in a jumble at the
location of the Virgo cluster.  M87 harbors in its center the largest black
hole yet discovered with a mass of $3 \times 10^9$ solar masses.

The dots appearing beyond M81 in the map are the 126,594 SDSS galaxies and
quasars (with $z < 5$) in the equatorial plane equatorial slice ($-2\degrees < \delta <
2\degrees$).  In addition, all 31 currently known SDSS quasars with $z > 5$ are
plotted, not just those
in the equatorial slice.  Since these large redshift quasars are shown from
all over the sky, a number of them appear at right ascensions which occur in the zone of avoidance
for the equatorial slice.  
The upper part of our map can be compared directly with figure~\ref{cpolbig}
and figure~\ref{cpolsmall}.  The map shows clearly and with recognizable
shape all the structures shown in the close-up in figure~\ref{cpolsmall},
while still showing all the SDSS quasars shown in the full view in
figure~\ref{cpolbig}.  The logarithmic scale captures both scales
beautifully.  On the left, (at about 1.5h and 120~Mpc) we can see clearly
the large circular void visible in figure~\ref{cpolsmall}.  To the right, at
RA of 9h-14h and at a distance of 215 -- 370 Mpc we can see a Sloan Great
Wall in the SDSS data, longer than the Great Wall of Geller and Huchra (the
CfA2 Great Wall).  The blank regions are where the Earth's equator cuts the
galactic plane and intersects the zone of avoidance near the galactic plane
(where the interstellar dust obscures distant galaxies and which the Sloan
survey does not cover).  These are empty fan-shaped regions as shown in
figure~\ref{cpolbig} and figure~\ref{cpolsmall}, bounded by radial lines
pointing away from the Earth, so on our map these are bounded by vertical
straight lines.

The Great Attractor (which is far off the equatorial plane and toward which the
Virgo super cluster has a measurable peculiar velocity) is shown.

The most prominent feature of the SDSS large scale structure seen in figure~\ref{p0} is
the ``Sloan Great Wall''. This feature was noticed early-on in the Sloan
data acquisition process, and has been mentioned in passing in a couple of
times in Sloan reports, accompanied by phrases such as ``large''
(\cite{bla03}), and ``striking'', ``wall-like'', and ``may be the largest
coherent structure yet observed'' (Tegmark et al., astro-ph/0310725). To
make a quantitative comparison,  we have also shown the Great Wall of Geller and Huchra.  This
extends over several slices of the CfA2 survey (from $42\degrees$ to $-8.5\degrees$
declination). Rather than plotting points for it here, which would be
confused with SDSS survey galaxies we have plotted density contours
averaging over all the CfA2 slices from $42\degrees$ to $-8.5\degrees$ declination.  This
volume extends far above the equatorial plane, and since we are plotting it
in right ascension correctly, it is not presented conformally, but is being
lengthened in the tangential direction by a factor of $\sim 1/\cos(21\degrees)
= 1.07$. Note that since the CfA2 Great Wall is a factor of approximately 2.5
closer to us than the Sloan Great Wall it is depicted at scale that
is 2.5 times larger.  So although the CfA2 Great Wall stretches from 9h to 16.7h
(or 7.7 hours of right ascension), as compared with the SDSS Great
Wall which stretches from 8.7h to 13.7h (or 5 hours of right
ascension) its real length in co-moving coordinates relative to the CfA2 Great Wall is, by this
simple analysis, $2.5 \times 5/(7.7 \times 1.07) \approx 1.74$ times as long. This is apparent in the
comparison figure supplied in figure~\ref{sloanwall}, where both are shown
at the same scale in co-moving coordinates. To make a fair comparison, since the Great Wall is almost
a factor of 3 closer than the Sloan Great Wall, we have plotted a
$12\degrees$ wide slice from the CfA2 survey to compare with our $4\degrees$
wide slice in the Sloan, so that both slices have approximately the same
width at each wall.

Of course, the walls are not perfectly aligned with the $x$ axis of our map,
so one has to measure their length along the curve. The Sloan Great Wall is
at a median distance of 310~Mpc. It's total length in co-moving coordinates
is 450~Mpc as compared with the total length of the Great Wall of Geller and
Huchra which is 240~Mpc long in co-moving coordinates. This indicates the
sizes the two walls would have at the current epoch. But the Great Wall is at a
median redshift of $z = 0.029$ so it's true size at the epoch we are
observing it is smaller by a factor of $1+z$ giving it an observed length of
232.64~Mpc (or 758 million light years). The Sloan great wall is at a
redshift of $z = 0.073$ so it's true observed length is
$419$~Mpc (or 1,365 million light years). For comparison, the CMB sphere has
an observed diameter of $(2 \cdot 14,000)/1090 = 25.7$~Mpc. The
observed length of the Sloan Great Wall is thus 80\% greater than the Great
Wall of Geller and Huchra.

Since we have numerous studies that show that the 3D topology of large scale
structure is sponglike (\cite{gdm86,vog94,hik02}) it should not be
surprising that as we look at larger samples we should find examples of
larger connected structures. Indeed, we would have had to have been
especially lucky to have discovered the largest structure in the observable
universe in the initial CfA survey which has a much smaller volume than the
Sloan survey. Simulated slices of the Sloan using flat lambda models (as
suggested by WMAP) show
great walls and great wall complexes that are quite impressive
(\cite{col00}). \cite{chw98} for example had a great wall in their
$\Omega_m = 0.4,\,\Omega_\Lambda = 0.6$ Sloan simulation which is 8\% longer
than the Great Wall of Geller and Huchra; and so it could be said that the
existence of a Great Wall in the Sloan longer than the Great Wall of Geller
and Huchra was predicted in advance. Visual inspection of the 275 PThalos
simulations reveals similar structures to the Sloan Great Wall in more than
10\% of the cases (Tegmark et al, astro-ph/0310725). Thus, it seems
reasonable that the Sloan Great Wall can be produced from random phase
Gaussian fluctuations in a standard flat-lambda model, a model that also
predicts a spongelike topology of high density regions in 3D. Notably, our
quantitative topology algorithm applied to the 2D Sloan Slice identifies the
Sloan Great Wall as one connected structure (\cite{hoy02}). Figure 2 in
\cite{hoy02} clearly shows this as one connected structure at the median
density contour when smoothed at $5h^{-1}$~Mpc in a volume limited sample
where the varying thickness and varying completeness of the survey in
different directions are accounted for. It is perhaps no
accident that both the Sloan Great Wall and the Great Wall of Geller and
Huchra are seen roughly tangential to the line of sight. 
Great Walls tangential to the line of sight are simply easier to see in slice
surveys, as pointed out by \cite{pra97}.
A Great Wall perpendicular to the line of sight would be more difficult to
see because the near end would be lost due to thinness of the slice and the
far end would be lost due to the lack of galaxies bright enough to be
visible at great distances. Redshift space distortions on large scales, i.e.
infall of galaxies from voids onto denser regions enhances contrast for real
features tangential to the line of sight. "Fingers of God" also make
tangential structures more noticable by thickening them. When \cite{par90} first simulated a volume large
enough to simulate the CfA survey, a Great Wall was immediately seen in the
3D data. When a slice to simulate the CfA slice seen from Earth -- which
gets wider as it gets further from the Earth -- was made, the Great Wall in
the simulation was pretty much a dead ringer for the Great Wall of Geller
and Huchra -- equal in length, shape, and density. This was an impressive
success for N-body simulations. The Sloan Great Wall and the CfA Great Wall
have been found in quite similar circumstances, each in a
slice of comparable thickness, and as illustrated in figure 9, both are
qualitatively quite similar except that the Sloan Great Wall is simply
larger, and as we have noted, our 2D topology algorithm (\cite{hoy02})
identifies the Sloan Great Wall as one connected structure in a volume
limited survey where the varying thickness and completeness of the slice
survey are properly accounted for. The CfA Great Wall is as large a structure 
as could have fit in the
CfA sample, but the Sloan Great Wall is smaller than the size of the Sloan
survey, showing the expected approach to homogeneity on the very largest
scales.

The 2dF survey (\cite{x2dF}), of similar depth to the Sloan, completed two slices, an
equatorial slice ($9^h:50^m < \alpha < 14^h:50^m, -7.5\degrees < \delta <
2.5\degrees$), and a southern slice ($21^h:40^m < \alpha < 03^h:40^m,
-37.5\degrees < \delta < -22.5\degrees$). This survey was thus not
appropriate for our logarithmic map of the universe. The southern slice was
not along a great circle in the sky, and therefore would be streched if
plotted in our map in right ascension. The equatorial slice was of less
angular extent than the corresponding Sloan Slice and so the Sloan with its
greater coverage in a flat equatorial slice was used to plot large scale
structure in our map. Indeed, the 2dF survey, because of its smaller
coverage in right ascension, missed the western end of the Sloan Great Wall
and so the wall did not show up as prominantly in the 2dF survey as in the
Sloan. Power spectrum analysis of the 2dF and the Sloan come up with quite
similar estimates. These two great surveys in many ways complement each
other. Perhaps most importantly the 2dF power spectrum analysis which was
available before Sloan and in time for WMAP allowed estimates of $\Omega_m$
which allowed WMAP to refine the cosmological parameters used in the
construction of this map. The Sloan Great Wall contains a number of Abell clusters (including,
for example, A1238, A1650, A1692 and A1750 for which redshifts are known).
The spongelike nature of 3D topology means that clusters are
connected by filaments or walls but if extended far enough, walls should
show holes allowing the voids on each side to communicate.

Indeed, in our map we can see some remnants of the CfA2 Great Wall (a couple of clumps or "legs")
extending into the equatorial plane of the Sloan sample.  As shown in
\cite{vog94}, if extended to the south the Great Wall develops
holes that allow the foreground and background voids to communicate leading
to a spongelike topology of the median density contour surface in 3D. In 3D,
the Sloan Great Wall may be connected to the supercluster of Abell
clusters found by Bahcall and Soneira (\cite{bs84}) whose two members lie just above it
in declination.

In the center of the Great Wall is the Coma Cluster, one of the largest
clusters of galaxies known.

The quasar 3C273 is shown as a cross. 

The gravitational lens quasar 0957 is shown as well as the lensing galaxy
producing the multiple image.  The lensing galaxy is along the same line of
sight but at about one third the co-moving distance.

The gamma ray burster GRB990123 is shown - for a brief period this was the
most luminous object in the observed universe.

The redshift $z = 0.76$ is shown as a line which marks the epoch that
divides the universe's decelerating and accelerating phase.  Objects closer
than this line are observed at an epoch when the universe's expansion is
accelerating, while objects further away than this line are observed at an
epoch when the universe's expansion is decelerating.

The unreachable limit is shown at $z = 1.69$.  Because of the acceleration
of the expansion of the universe, photons sent from Earth now will never
reach objects beyond this line.  Galaxies beyond this line will never hear
our current TV signals. Spaceships from Earth, traveling slower than light
will also find the territory beyond this line unreachable.  This redshift is
surprisingly low.  It is interesting that we can see many objects today that
are so far away that we can never get to them.
 
SDSS quasars in the equatorial plane ($-2\degrees < \delta < 2\degrees$) are shown as points out
to a redshift of $z = 5.0$ using redshifts determined from the SDSS survey spectra.  For
quasars with $z > 5.0$ we have shown all 31 quasars from the SDSS with $z > 5$
regardless of declination.  There are now a surprisingly large number of
quasars with $z > 5$ known.  Because a number of these are at high
declination, they occur at right ascensions that are in the zone of avoidance for the equatorial
plane.  Each is shown at its proper right ascension and distance, including
the largest redshift one with $z = 6.42$ which is labeled (M. Strauss
2003, private communication).  This is currently the largest redshift quasar known.

The galaxy SDF~J132418.3+271455 with the largest accurately measured
redshift ($z = 6.578$) is also shown. This was discovered in the Subaru Deep
Field (\cite{kod03}).

The co-moving distance at which the first stars are expected to form is
shown as a dashed line.  The WMAP satellite has found that the first stars
appear about 200 million years after the Big Bang and the map therefore
indicates the distance out to which stars could be seen in principle.

Finally there is the cosmic microwave background radiation discovered by
Penzias and Wilson in 1965 (\cite{pw65}).  CMB photons from this surface arrive directly
from an epoch only 380,000 years after the Big Bang.

A line showing the co-moving radius back to the Big Bang is also shown. 
This represents seeing back to the epoch just after inflation. 
The co-moving distance between the cosmic
microwave background and the Big Bang is shown in correct proportion to the
circumference.  In comparing with figure~\ref{cpolbig}, we note that in that map
the circumference is a factor of $\pi$ larger.  So in our map, which shows
the $360\degrees$ circumference of the cosmic microwave background as
approximately the same length as the diameter of the circle in figure
\ref{cpolbig}, the scale at that point is about a factor of
$\pi$ smaller than in figure~\ref{cpolbig} and the cosmic microwave
background and the Big Bang are closer to each other (by a factor of $\pi$)
than in figure~\ref{cpolbig} as expected.

Last is shown the co-moving future visibility limit.  If we wait until the
infinite future we will eventually be able to see the Big Bang at the
co-moving future visibility limit.  Stars and galaxies that lie beyond this
co-moving future visibility limit are forever hidden from our view.  Because
of the de\ Sitter expansion produced by the cosmological constant the universe
has an event horizon which we cannot see over no matter how long we wait.

It is remarkable how many of the features shown in this map have been
discovered in the current astronomical generation.  When one of us (JRG)
began studing astronomy at age 8 (in 1955), on astronomical maps there were
no artificial satellites, no Kuiper belt objects, no other stars with
planets, no brown dwarfs, no pulsars, no black holes, no non-solar X-ray sources, no gamma ray
bursters, no great walls, no great attractors, no quasars, no gravitational
lenses, and no observation of the cosmic microwave background.

\section{Applications}

This map shows large scale structure well. "Fingers of God" are vertical
which makes removing them for large scale structure purposes particularly
easy. A test for roundness of voids as proposed by \cite{ryd95} -- the
Alcock-Paczynski isotropy test -- could be done on this map as well as on the
co-moving map.  The ability of this test to differentiate between
cosmological models can be deduced by measuring void isotropy as a function
of cosmological model in the plane of parameters ($\Omega_m$, $\Omega_\Lambda$).  For the correct
cosmological model the void pictures will be isotropic (since the Earth is
not in a special position in the universe).  To do this test we need
conformal maps for various cosmological models, and so we need conformal
maps for the $k = +1$ and $k = -1$ cases as well as the $k = 0$ case so that
statistical comparisons can be made.  The formulas for these projections are
given in the appendix.

A Fourier analysis of large scale structure modes in the map $(k_x, k_y)$ can
provide information on the parameter $\beta = \Omega_m^{0.6}/b$ where $b$ is the bias
parameter.  For
fluctuations in the linear regime in redshift space $d\rho/\rho \propto (1 + \beta \mu^2)$ where $\mu
= \cos\theta$ where $\theta$ is the angle between the normal to the wave and the line of
sight in 3D (\cite{kai87}).  Waves tangential to the line of sight
have a larger amplitude in redshift space than waves parallel to the line of
sight because the peculiar velocities induced by the wave enhance the
amplitude of the wave when peculiar radial velocities are added to Hubble
positions as occurs when galaxies are plotted using redshift at the Hubble
flow positions assuming peculiar velocities are zero. Imagine a series of
waves isotropic in 3D. Then one
can show that the average value of $\mu^2$ in 3D of those waves with an
orientation $\mu'=\cos(\phi)$ in the equatorial slice is $\langle \mu^2 \rangle =
(2/3) \mu'^2$. Thus, we expect approximately, for waves observed in our equatorial slice
$\delta\rho/\rho \propto (1 + [2/3]\beta\mu'^2)$.  Waves in our map with constant $(k_x, k_y)$ represent
global logarithmic spiral modes with constant inclination relative to the
line of sight.  Of course this is an oversimplified treatment, since we must
consider the power spectrum of fluctuations when relating the modes seen at
a particular wavelength in the plane and in 3D and the "fingers of God" must
be eliminated by some friend-of-friend algorithm.  But in general, we expect
that modes that are radial ($k_x = 0$) will have higher amplitudes than modes
that are tangential ($k_y = 0$) and this effect can be used empirically, in
conjunction with N-body simulations, to provide an independent check on the
value of $\beta$.  One simply adopts a cosmological model, checks with N-body
simulations the relative amplitude of map modes as a function of $\mu'$ in the
logarithmic map and compares with the observations assuming the same
cosmological model when plotting the logarithmic map: if the cosmological
model $(\Omega_m, \Omega_\Lambda)$ is correct, the results should be similar.

Since logarithmic spirals appear as straight lines in our map projection, it
may prove useful for mapping spiral galaxies.  Take a photograph of a
face-on spiral galaxy, place the origin of the coordinate system in the
center of the galaxy and then construct our logarithmic map of this 2D
planar photograph.  The spiral arms (which approximate logarithmic spirals)
should then appear as straight lines on the conformal map.  We have tried
this on a face-on spiral galaxy photograph given to us by James Rhoads and
the results were very satisfying.  The spiral arms indeed were beautiful
straight lines, and from their slope one could easily measure their inclination
angle.  Star images in the picture were still circular in the map because
the map is conformal.

\section{Conclusions}

Maps can change the way we look at the world.  Mercator's map presented in
1569 was influential not only because it was a projection that showed the
shapes of continents well but because for the first time we had pretty
accurate contours for North America and South America to show.  The
\underline{Cosmic View} and the \underline{Powers of Ten} alerted people to the scales in the universe we
had begun to understand.  De Lapparent, Geller, and Huchra showed how a slice of the
universe could give us an enlightening view of the universe in depth.  Now
that astronomers have arrived at a new understanding of the of the universe
from the solar system to the cosmic microwave background, we hope our map
will provide in some small way a new visual perspective on these exciting
discoveries.

The map presented here is appropriate for use as a wall map. It's scale
is approximately 1~inch/radian. A version at twice the scale, 12.56~inches
by 95.2~inches tall would also be appropriate for a wall chart and
would run nearly from floor to ceiling in a normal room with an 8ft ceiling. 
If one wanted to show individual objects at $60\times$ scale, the Moon and Sun
would be 1~inch across, M31 would be 4~inches across, and Mars would be
0.0145~inches in diameter and Jupiter 0.0272~inches. Alternately the Sun and
Moon and Messier objects could be shown a $60\times$ scale with the planets at $600\times$ scale to
illustrate their usual appearance in small telescopes.

Consider some possible (and some fanciful) ways this map of the universe
might be presented for educational use.

We have presented the map on the internet in color on astro-ph.  In
principle it would be easy to have such a map on the internet automatically
continuously updated to track the current positions of the satellites, Moon,
Sun, asteroids, planets, and Kuiper belt objects as a function of time, in
fact we have plotted them as of a particular date and time using such
programs.  New objects could be added to the map as they were discovered. 
Click on an object, and a 60x enlarged view of it would appear.  Two clicks,
and a 3600x enlarged view would appear, and so forth -- until the highest
resolution picture available was presented.  Individual images of all the
SDSS galaxies and quasars shown on the map could be accessed in this way, as
well as M objects .  When an individual object was selected, helpful
internet links to sites telling more about it would appear.

Since the left and right hand edges of the map are identical, the map could
be profitably shown as a cylinder. In cylindrical form, the map at the approximate scale
and detail of figure 7 is in perfect proportion to be used on a pencil with
the Earth's surface at the eraser end and the Big Bang at the point end.  
Perhaps the
best cylindrical form for the map would be a cylinder on the interior of an
elevator shaft for a glass elevator. Every floor you went up you
would be looking at objects that were 10 times further away than the preceding
floor.  A trip up such an elevator shaft could be simulated in a planetarium show, with the
cylindrical map being projected onto the dome showing the view from the
elevator as it rose.
  
We have put our map up on Princeton's flat video wall. 
This wall has a horizontal resolution of 4096 pixels.  From
top to bottom of the video wall is about 1600 pixels, so the scale change in
the map from top to bottom is about a factor of 10.  This shows a small
portion of the entire map.  We then scan this in real time moving
steadily upward from the Earth to the cosmic microwave background and the
Big Bang. This produces
a virtual map 17.6 feet wide and 134 feet tall.  With laser beams it would be
easy to paint a large version of the map on the side of a building. A very large 
temporary version of our map of the universe could also be set up in a park 
or at a star party by simply planting markers for the salient objects.

The map could be produced on a carpet 6 feet wide 45.5 feet long for a
hallway in an astronomy department or planetarium.  Then every step you took down the
hallway would take you about a factor of about ten further from the
Earth -- a nice way to have a walk through the Universe.

\acknowledgments

    We would like to thank Michael Strauss for supplying us with the list of
SDSS quasars with redshift greater than 5.

    This work was supported by JRG's NSF grant AST04-06713.


    Funding for the creation and distribution of the SDSS Archive has been
provided by the Alfred P. Sloan Foundation, the Participating Institutions,
the National Aeronautics and Space Administration, the National Science
Foundation, the U.S. Department of Energy, the Japanese Monbukagakusho, and
the Max Planck Society. The SDSS Web site is http://www.sdss.org/.

    The SDSS is managed by the Astrophysical Research Consortium (ARC) for
the Participating Institutions. The Participating Institutions are The
University of Chicago, Fermilab, the Institute for Advanced Study, the Japan
Participation Group, The Johns Hopkins University, the Korean Scientist
Group, Los Alamos National Laboratory, the Max-Planck-Institute for
Astronomy (MPIA), the Max-Planck-Institute for Astrophysics (MPA), New
Mexico State University, University of Pittsburgh, Princeton University, the
United States Naval Observatory, and the University of Washington.

\appendix

\section{$k = +1$, $k = 0$ and $k = -1$ cases}

Although the observations suggest the $k = 0$ case is appropriate for the
universe, for mathematical completeness we consider the general Friedmann
metrics:
\begin{eqnarray}
ds^2 & = & -dt^2 + a^2(t)[d\chi^2 + \sin^2\chi(d\theta^2 + \sin^2\theta\cdot d\phi^2)], \quad k = +1, \label{closedmetric} \\
ds^2 & = & -dt^2 + a^2(t)[d\chi^2 + \chi^2(d\theta^2 + \sin^2\theta\cdot d\phi^2)], \quad k = 0, \label{flatmetric} \\
ds^2 & = & -dt^2 + a^2(t)[d\chi^2 + \sinh^2\chi(d\theta^2 + \sin^2\theta\cdot d\phi^2)], \quad k = -1 \label{openmetric}
\end{eqnarray}

Define the conformal time by:
\begin{equation}
\eta(t) = \int_0^t\frac{d t}{a} = \int_0^{a(t)} (-k a^2 + \frac{8\pi}{3}a^4\{\rho_m(a) + \rho_r(a)\} + \frac{\Lambda}{3}a^4)^{-1/2} da
\end{equation}
If we are currently at the epoch $t_0$, then the current conformal time is
$\eta(t_0)$.  When we look back to a redshift $z$ we are seeing an epoch when
$a_0/a(t) = 1 + z$, and out to a co-moving distance $\chi$ given by:
\begin{equation}
\chi(z) = \eta(t_0) - \eta(t) = \int_{\frac{a(t_0)}{1+z}}^{a(t_0)}
	(-k a^2 + \frac{8\pi}{3}a^4\{\rho_m(a) + \rho_r(a)\} + \frac{\Lambda}{3}a^4)^{-1/2} da
\end{equation}
As we have noted, in the flat case $k = 0$, we are free to adopt a scale for
$a$, so we set $a(t_0) = R_{H_0}$.

For the $k = +1$ case, a two dimensional slice through the universe is a
sphere.  So we need to make a conformal projection of the sphere onto a
plane, and then we can apply our conformal logarithmic projection as before
to produce our map. The stereographic map projection is such a conformal
projection of a sphere onto a plane.  Adopt coordinates on the sphere of $\chi$,
$\theta$, where the metric on the sphere is given by:
\begin{equation}
ds^2 = a^2(t_0)(d\chi^2 + \sin^2\chi\cdot d\theta^2)
\end{equation}
Where the angle $\theta$ is now a longitude (Thus, $\theta$ is equivalent to $\phi$ in metric
\ref{flatmetric} above -- where in metric \ref{flatmetric} we are considering the equatorial slice with $\theta =
const. = \pi/2$. So to get the metric above from metric \ref{flatmetric} we set $\theta = const. =
\pi/2$, and replace $\phi$ with $\theta$).  Now make a stereographic conformal projection
of this sphere $(\chi, \theta)$ onto a plane with polar coordinates $(r,\theta)$:
\begin{eqnarray}
r      & = & 2a(t_0)\tan(\frac{\chi}{2}) \\
\theta & = & \theta
\end{eqnarray}
This is a projection from the north pole onto a plane tangent to the south
pole.  The Earth would be at the south pole of the sphere (where $\chi = 0$).  A
line from the north pole to the plane through the point $(\chi, \theta)$ on the sphere
will be at an angle $\alpha = \chi/2$ relative to a normal to the plane, and also at
an angle $\alpha = \chi/2$ relative to a normal to the surface of the sphere at the
point $(\chi,\theta)$, thus the foreshortening that occurs along the ray from the
north pole as it crosses the surface of the sphere is exactly the same as
the foreshortening that occurs when it crosses the surface of the plane. 
Thus, shapes in the surface are locally mapped without distortion locally
onto the plane.  The map is conformal.  Now we have a conformal map of the
sphere and we apply our logarithmic conformal map to the planar map to get
our map of the universe:
\begin{eqnarray}
x & = & -\theta \\
y & = & \ln\frac{r}{r_E} = \ln \frac{2a(t_0)\tan(\frac{\chi}{2})}{r_E} \label{xyopen}
\end{eqnarray}
This provides a conformal mapping from $(\chi, \theta)$ to $(x, y)$.  It is
in fact a Mercator projection of the spherical section of the universe with
the Earth located at the south pole of the sphere!  We are used to seeing
the Mercator projection for the surface of the Earth cut off at the
Antarctic circle, so we do not see that if it were extended much nearer to
the south pole (in the limit as $\chi \to 0$) it would approximate our
logarithmic map of a plane in the region of the south pole. 
(This projection would be useful for cosmological models that were slightly closed.)

For the $k = -1$ case, a 2D slice of the universe is a negatively curved
pseudosphere with metric:
\begin{equation}
d s^2 = a^2(t_0)(d\chi^2 + \sinh^2\chi(d\theta^2)), \quad k = -1
\end{equation}
Where $\chi$ is the co-moving radius, and as before we have set $\theta = const. = \pi/2$,
in metric \ref{openmetric} above to look at an equatorial slice, and we have replaced $\phi$
in metric \ref{openmetric} with $\theta$, so that $\theta$ is now a longitude.  We can conformally
project the pseudosphere onto plane with a map projection that is an
analogue of the stereographic projection for the sphere.  The metric above
for a pseudosphere of radius $a(t_0)$ is the metric on the hyperboloid
surface:
\begin{equation}
x^2 + y^2 - t^2 = -a^2(t_0)
\end{equation}
where $t > 0$ in a three dimensional Minkowski space with metric:
\begin{equation}
ds^2 = - dt^2 + dx^2 + dy^2
\end{equation}
define coordinates on this surface $(\chi, \theta)$ by 
\begin{eqnarray}
	t & = & a(t_0) \cosh \chi		\\
	y & = & a(t_0) \sinh \chi \sin \theta	\\
	x & = & a(t_0) \sinh \chi \cos \theta
\end{eqnarray}
with the definitions above it is easy to see that $x^2 + y^2 - t^2 = -a^2(t_0)$,
since $a^2(t_0)\sinh^2\chi \cos^2\theta + a^2(t_0)\sinh^2\chi \sin^2\theta - a^2(t_0)\cosh^2\chi = - a^2(t_0)$.
Connect each point $(\chi, \theta)$ on the surface with the origin $(x, y, t) = (0, 0, 0)$
via a straight worldline.  This worldline has a velocity relative to the $t$
axis of $v = \sinh\chi/\cosh\chi = \tanh{\chi}$.  So this worldline has a boost of $\chi$
relative to the $t$ axis.  This worldline is normal to the surface at the
point $(\chi, \theta)$, because that boost takes the $t$ axis to the worldline in
question and leaves the hyperbolic surface invariant.  Let the intersection
of that worldline with the tangent plane $t = a(t_0)$ be the gnomonic map
projection of the point $(\chi, \theta)$ onto the plane with polar coordinates $(r,\theta)$.
Then
\begin{eqnarray}
	r      & = & a(t_0) \tanh \chi \\
	\theta & = & \theta
\end{eqnarray}
This gnomonic projection maps the pseudosphere $0 \le \chi \le \infty$ into a disk of
radius $r = a(t_0)$.  Why?  Because in a time $t = a(t_0)$ the worldline which has a
velocity $v = \tanh \chi$ travels a distance of $r = vt = a(t_0) \tanh \chi$.  This
gnomonic projection is not conformal, because the worldline is normal to the
hyperboloid surface but not normal to the plane $t = a(t_0)$. This gnomonic
projection maps geodesics on the negatively curved hyperboloid onto
straight lines on the plane $t = a(t_0)$ because any plane in the space $(x, y, t)$ passing
through the origin intersects the surface in a geodesic and this plane will
intersect the tangent plane $t = a(t_0)$ in a straight line.

A conformal projection, like the stereographic projection of the sphere is
provided for the pseudosphere by connecting with a worldline each point on the surface $(\chi, \theta)$,
to the point $(x, y, t) = (0, 0, -a(t_0))$, and letting this worldline intersect
the plane $t = a(t_0)$ at a point $(r, \theta)$.  Now that worldline has a velocity 
$v = \sinh\chi/(\cosh\chi + 1) = \tanh (\chi/2)$ (using a half angle trigonometric
identity).  Thus, this worldline has a boost relative to the $t$ axis of $\chi/2$
and therefore a boost of $\chi/2$ relative to the normal to the plane $t = a(t_0)$.
When it intersects the hyperbolic surface at the point $(\chi, \theta)$, the normal to the hyperbolic surface
at that point has a boost of $\chi$ relative to the $t$ axis in the same plane.
Thus, the worldline has a boost of $\chi/2$ relative to the normal of the
tangent plane and also a boost of $\chi/2$ relative to the hyperbolic surface,
so it has an equal boost (and observes equal Lorentz contractions) relative
to both (giving identical foreshortenings as in the stereographic projection
of the sphere) and therefore the map projection is conformal.  This
worldline connecting the point $(x, y, t) = (0, 0, -a(t_0))$ to the point $(\chi, \theta)$ on
the surface thus intersects the tangent plane $t = a(t_0)$ at the point $(r, \theta)$
given by
\begin{eqnarray}
     r & = & 2a(t_0) \tanh (\frac{\chi}{2}) \\
\theta & = & \theta
\end{eqnarray}
This maps the pseudosphere $0 \le \chi < \infty$ into a disk of radius $r = 2a(t_0)$.  Now
we have a conformal map projection of the negatively curved $k = -1$ universe
onto a plane.  This conformal map is the one illustrated in Escher's famous
angels and devils print which is often used to illustrate this cosmology
(cf. \cite{got01}). Next we take this conformal planar map $(r, \theta)$ and apply
our conformal logarithmic mapping function to it.  Thus, we produce a
conformal map of the universe with coordinates:
\begin{eqnarray}
	x & = & -\theta	\\
	y & = & \ln\frac{2a(t_0)\tanh(\chi/2)}{r_E}
\end{eqnarray}
Note the similarity with the formula for a spherical $k = +1$ universe; in the 
$k = -1$ case ``$\tanh$'' simply replaces ``$\tan$''.

We can make maps for all three cases.  These are useful for comparison
purposes.  For example suppose we develop an isotropy measure for voids as
suggested by Barbara Ryden.  We could apply this to the voids on our
conformal universe map since it preserves shapes locally and the voids are
small.  The shapes of the voids depend on the assumed cosmological model. 
This is the Alcock--Paczynski test.  If we have the right cosmological
model the voids will be approximately round as suggested by Ryden, and this
can be tested.  Suppose, as WMAP suggests, the flat $k = 0$ case is correct
with $\Omega_m + \Omega_\Lambda = 1$, Then we can plot the same redshift data but conformally
assuming slightly closed $\Omega_m + \Omega_\Lambda > 1$, $k = +1$ models or slightly 
open $\Omega_m + \Omega_\Lambda < 1$, $k = -1$ models and check isotropy of the voids in each case.  We
can thus check how sensitive the isotropy test is in limiting the location
of the model in the $(\Omega_m, \Omega_\Lambda)$ parameter plane.

\section{Sources}

In applying the logarithmic map to the problem of showing the Universe, we have used a
multitude of both online sources, personal communication and data published
in journals. To illustrate the wealth and variety of data depicted in the
Map, we have chosen to list the sources in this section.

\begin{itemize}
	\item Moon phase \\
		United States Naval Observatory Astronomical Almanac,
		(\cite{usnoalmanac}) \\
		( http://aa.usno.navy.mil/data/docs/MoonPhase.html )

	\item Earth geological data \\
		Allen's Astrophysical Quantities (\cite{cox00})

	\item Artificial satellites \\
		Mike McCants' Satellite Tracking web pages
		  (alldat.tle file). The file includes orbital elements supplied
		  by OIG (NASA/GSFC Orbital Information Group) and data on
		  other satellites obtained primarily by amateur visual
		  observations. \\
		( http://oig1.gsfc.nasa.gov/ ) \\
		( http://users2.ev1.net/\~{}mmccants/tles/index.html )

	\item Hubble Space Telescope location \\
		Alan Patterson \& Davin Workman (Science and Mission
		Scheduling Branch, Operations and Data Management Division,
		Space Telescope Science Institute), private communication

	\item Chandra X-ray Observatory location \\
		Rob Cameron (CXO Science Operations Team), private communication

	\item Asteroid catalog \\
		Lowell Observatory Asteroid Database (ASTORB), 2003/04/19 snapshot, with ephemeris calculated for Aug 12, 2003 \\
		ftp://ftp.lowell.edu/pub/elgb/astorb.html

	\item Minor bodies' ephemeris \\
		Ephemeris computed using an adapted version of the OrbFit software package. \\
		OrbFit is written by the OrbFit consortium: Dept. of
	          Mathematics, Univ. of Pisa (Andrea Milani, Steven R.
	          Chesley), Astronomical Observatory of Brera, Milan (Mario
	          Carpino), Astronomical Observatory, Belgrade (Zoran
	          Kne\v{z}evi\'c), CNR Institute for Space Astrophysics, Rome
	          (Giovanni B.  Valsecchi). \\
		( http://newton.dm.unipi.it/\~{}neodys/astinfo/orbfit/ )

	\item Quaoar and Sedna information \\
		( http://www.gps.caltech.edu/\~{}chad )

	\item Comet, Sun, Moon and planet ephemeris \\
		Generated using JPL HORIZONS On-line Solar System Data and Ephemeris Computation Service. \\
		( http://ssd.jpl.nasa.gov/horizons.html )

	\item Space probes \\
		Voyager 1, Voyager 2, Pioneer 10, Ulysses positions obtained
		from NASA HeloWeb webpage at NSSDC (National Space Science
		Data Center) \\
		( http://nssdc.gsfc.nasa.gov/space/helios/heli.html )

	\item Heliopause \\
		Distance taken to be approximately 110~AU in the direction of approximately 18h RA.

	\item Oort Cloud \\
		Taken to extend from 8,000~AU to 100,000~AU

	\item 10 brightest stars \\
		Source for positions and parallaxes: SIMBAD Reference database, Centre de Donnees astronomiques de Strasbourg \\
		( http://simbad.u-strasbg.fr/sim-fid.pl )

	\item 10 nearest stars \\
		Source for positions and parallaxes: Research Consortium on Nearby Stars, list as of July 1, 2004 \\
		( http://www.chara.gsu.edu/RECONS/TOP100.htm )

	\item Extra-solar planets \\
		IAU ``Working Group on Extrasolar Planets'' list of planets,
		  the "Extrasolar Planets Catalog" of the "Extrasolar
		  Planets Encyclopedia" maintained by Jean Schneider at CNRS
		  -- Paris Observatory \\
		( http://www.ciw.deu/IAU/div3/wgesp/planets.shtml )\\
		( http://www.obspm.fr/encycl/encycl.html )\\
		( http://cfa-www.harvard.edu/planets/OGLE-TR-56.html )

	\item Hipparcos stars \\
	        ESA, 1997, The Hipparcos and Tycho Catalogues, ESA SP-1200 \\
		( http://tdc-www.harvard.edu/software/catalogs/hipparcos.html )

	\item Selected Messier objects \\
		Distances and common names taken from the Students for the
		  Exploration and Development of Space (SEDS) Messier
		  Catalog pages. Positions taken from the SIMBAD reference
		  database \\
		( http://www.seds.org/messier/ )

	\item Milky Way \\
		Disk radius taken to be 15kpc.  Distance to galactic center taken to be 8kpc.

	\item Local Group \\
		Data taken from a list of Local Group Member Galaxies maintained by SEDS \\
		( http://www.seds.org/\~{}spider/spider/LG/lg.html )

	\item Great Attractor \\
		Location data from SIMBAD, redshift distance: $cz = 4,350\,\mathrm{km\,s^{-1}}$

	\item Great Wall \\
		Contours based on CfA2 redshift Catalog, subset CfA2. The contours are based on galaxies
		  satisfying $-8.5\degrees < \delta < 42.5\degrees$, $120\degrees < \alpha < 255\degrees$,
                  and $0.01 < z < 0.05$ comprising the CfA2's first 6 slices (\cite{gh89}). \\
		( http://cfa-www.harvard.edu/\~{}huchra )

	\item SDSS data \\
		Plotted from raw SDSS spectroscopy data obtained using David
		  Schlegel's SPECTRO pipeline -- spALL.dat file -- dated 2003/01/15 with a
		  $z < 5$ inclusion cut applied. \\
		SDSS quasar data with $z > 5$ provided by Michael Strauss, private communication. \\
		( http://spectro.princeton.edu/ )

	\item Plotting \\
		Plotted using SM software by Robert H. Lupton. \\
		( http://astro.princeton.edu/\~{}rhl/sm )

	\item Individual quasar data \\
		SIMBAD database

	\item Cosmological data \\
		WMAP Collaboration publications ( \cite{wmap} )

	\item WMAP location in space at the time of the map \\
		Dale Fink, Gary Hinshaw, Hiranya Peiris (2003), personal communication

\end{itemize}

\clearpage

\begin{figure}
\centering
\caption{Map of the Universe. This figure is available separately from
http://www.astro.princeton.edu/\~{}mjuric/universe\label{p0}}
\end{figure}

\begin{figure}
\centering
\includegraphics[width=\textwidth]{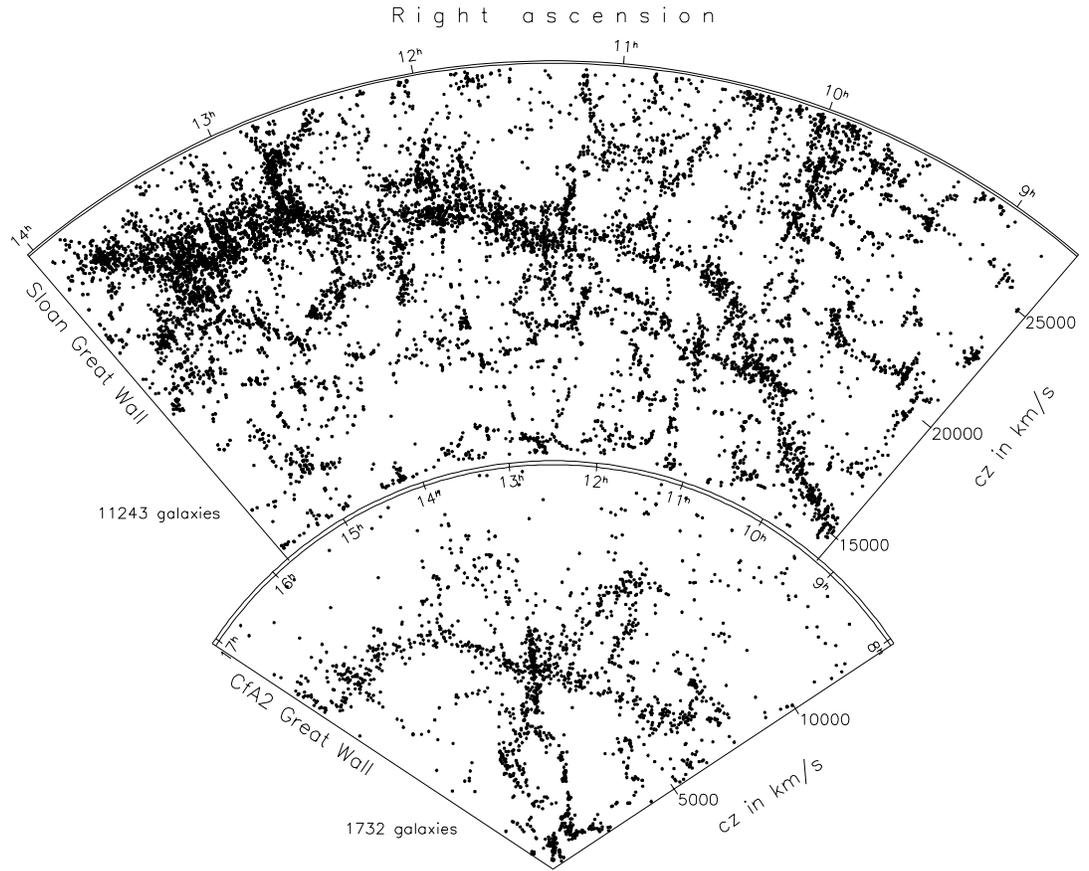}
\caption{Sloan Great Wall compared to CfA2 Great Wall at the same scale in co-moving coordinates. Equivalent
redshift distances $cz$ are indicated. The Sloan slice is $4\degrees$ wide,
the CfA2 slice is $12\degrees$ wide to make both slices approximately the
same physical width at the two walls. The Sloan Great Wall extends from 14h
to 9h. It consists of one strand at the left, which divides to form two
strands between 11.3h and 9.8h, which come back together to fom one strand
again (like a road that becomes a divided highway for a while). The CfA2
Great Wall (which includes the Coma cluster in the center), has been plotted
on a cone and then flattened onto a plane. Total numbers of galaxies shown
in each slice are also indicated.
\label{sloanwall}
}
\end{figure}

\end{document}